\documentclass[letterpaper,12pt]{article}

\usepackage[margin=1in]{geometry}
\usepackage{url}

\usepackage{amssymb}
\usepackage{amsmath}
\usepackage{amsthm}
\usepackage{xspace}
\usepackage{paralist}
\usepackage{verbatim}
\usepackage[linesnumbered,ruled,vlined]{algorithm2e}
\usepackage{graphicx}
\usepackage{subfig}
\usepackage{color}
\usepackage[all]{xy}

\newcommand{\Dfn}[1]{\textbf{\emph{#1}}}
\newcommand{\Inv}[1]{\textbf{Inv-#1}}
\newcommand{\Obj}[1]{\textbf{O#1}}
\newcommand{\ToDo}[1]{\emph{[\textbf{To Do:} #1]}}

\newcommand{\Nat}{\ensuremath{\mathbb{N}}\xspace}
\newcommand{\PartFunc}{\ensuremath{\not\rightarrow}\xspace}
\newcommand{\Dom}[1]{\ensuremath{\mathit{dom}(#1)}}
\newcommand{\Ran}[1]{\ensuremath{\mathit{ran}(#1)}}

\newcommand{\MapEx}[1]{\ensuremath{\mathit{ex}[#1]}}
\newcommand{\MapSerial}[1]{\ensuremath{\mathit{serial}[#1]}}
\newcommand{\MapState}[1]{\ensuremath{\mathit{state}[#1]}}
\newcommand{\MapMonitor}[1]{\ensuremath{\mathit{monitor}[#1]}}
\newcommand{\MapFragment}[2]{\ensuremath{\mathit{fragment}[#1, #2]}}
\newcommand{\MapBaton}[1]{\ensuremath{\mathit{baton}[#1]}}

\newcommand{\anSA}{\ensuremath{M}\xspace}
\newcommand{\SAAlpha}{\ensuremath{\Sigma}\xspace}
\newcommand{\SAStates}{\ensuremath{Q}\xspace}
\newcommand{\SAInit}{\ensuremath{q_0}\xspace}
\newcommand{\SATrans}{\ensuremath{\delta}\xspace}
\newcommand{\SATransEx}{\ensuremath{\delta^*}\xspace}
\newcommand{\SA}[4]{\ensuremath{( #1, #2, #3, #4 )}}
\newcommand{\anSAState}{\ensuremath{q}\xspace}

\newcommand{\SAStationary}[2]{\ensuremath{\mathsf{stat}_{#1}(#2)}}
\newcommand{\SATransitioning}[2]{\ensuremath{\mathsf{trans}_{#1}(#2)}}

\newcommand{\Pre}{\ensuremath{\mathit{prefixes}}\xspace}

\newcommand{\opPar}{\ensuremath{\operatorname{|\!|}}\xspace}
\newcommand{\opXor}{\ensuremath{\otimes}\xspace}
\newcommand{\opSeq}{\ensuremath{\operatorname{;}}\xspace}
\newcommand{\opInt}{\ensuremath{\cap}\xspace}

\newcommand{\aPerm}{\ensuremath{p}\xspace}

\newcommand{\SP}{\ensuremath{\mathit{SP}}\xspace}

\newcommand{\aPState}{\ensuremath{\gamma}\xspace}
\newcommand{\PStates}[1]{\ensuremath{\Gamma(#1)}\xspace}
\newcommand{\anAState}{\ensuremath{A}\xspace}
\newcommand{\anRState}{\ensuremath{R}\xspace}
\newcommand{\aCState}{\ensuremath{C}\xspace}
\newcommand{\PState}[4]{\ensuremath{( #1, #2, #3, #4 )}}
\newcommand{\AState}[2]{\ensuremath{( #1, #2 )}}
\newcommand{\RState}[2]{\ensuremath{( #1, #2 )}}
\newcommand{\anExR}{\ensuremath{\anEx_{\mathit{rs}}}\xspace}
\newcommand{\anSAStateA}{\ensuremath{\anSAState_{\mathit{as}}}\xspace}
\newcommand{\TSState}{\ensuremath{\aTS_{\mathit{as}}}\xspace}
\newcommand{\TSValid}{\ensuremath{\aTS_{\mathit{rs}}}\xspace}

\newcommand{\aTS}{\ensuremath{t}\xspace}
\newcommand{\TSClock}{\ensuremath{\aTS_{\mathit{clo}}}\xspace}

\newcommand{\TSSerial}{\ensuremath{\aTS_{\mathit{ser}}}\xspace}

\newcommand{\TSFirst}[1]{\ensuremath{\mathit{first}(#1)}}
\newcommand{\TSLast}[1]{\ensuremath{\mathit{last}(#1)}}
\newcommand{\Times}[1]{\ensuremath{\mathit{times}(#1)}}

\newcommand{\anEx}{\ensuremath{e}\xspace}
\newcommand{\ExNil}[1]{\ensuremath{\mathsf{nil}(#1)}}
\newcommand{\ExEx}[3]{\ensuremath{\mathsf{ex}(#1, #2, #3)}}

\newcommand{\anAssertion}{\ensuremath{\mathit{\alpha}}\xspace}
\newcommand{\aTicket}{\ensuremath{\mathit{tic}}\xspace}
\newcommand{\CCap}[2]{\ensuremath{\mathsf{cap}(#1, #2)}}
\newcommand{\CUpdate}[1]{\ensuremath{\mathsf{upd}(#1)}}
\newcommand{\aTicketSet}{\ensuremath{\mathit{Ts}}\xspace}

\newcommand{\SAFrag}[2]{\ensuremath{(#1, #2)}}
\newcommand{\Names}{\ensuremath{\mathcal{N}}\xspace}
\newcommand{\aName}{\ensuremath{n}\xspace}
\newcommand{\aCurrName}{\ensuremath{n_\star}\xspace}
\newcommand{\Unknown}{\ensuremath{\circ}\xspace}
\newcommand{\anSAFrag}{\ensuremath{F}\xspace}
\newcommand{\FTrans}{\ensuremath{\mathit{trans}}\xspace}
\newcommand{\FDefs}{\ensuremath{\mathit{defs}}\xspace}

\newcommand{\FDo}[2]{\ensuremath{\Delta(#1, #2)}}
\newcommand{\FDoEx}[2]{\ensuremath{\Delta^*(#1, #2)}}
\newcommand{\FDoExUpto}[3]{\ensuremath{\Delta^*_{\leq #1}(#2, #3)}}
\newcommand{\FDoExAfter}[3]{\ensuremath{\Delta^*_{> #1}(#2, #3)}}

\newcommand{\aKey}{\ensuremath{k}\xspace}
\newcommand{\Encap}[3]{\ensuremath{\langle\!\langle #3
    \rangle\!\rangle_{#1, #2}}}
\newcommand{\EncapCap}[5]{\Encap{#1}{#2}{#3: \CCap{ #4 }{ #5 }}}
\newcommand{\EncapUpdate}[4]{\Encap{#1}{#2}{#3: \CUpdate{ #4 }}}
\newcommand{\aSessID}{\ensuremath{\mathit{sessid}}\xspace}
\newcommand{\aUID}{\ensuremath{\mathit{uid}}\xspace}
\newcommand{\aCap}{\ensuremath{\mathit{cap}}\xspace}

\newcommand{\EncapCapMulti}[6]{\Encap{#1}{#2}{#3, #4: \CCap{ #5 }{ #6 }}}
\newcommand{\EncapUpdateMulti}[5]{\Encap{#1}{#2}{#3, #4: \CUpdate{ #5 }}}
\newcommand{\Validator}{\ensuremath{\mathit{vid}}\xspace}
\newcommand{\RSID}{\ensuremath{\mathit{rsid}}\xspace}
\newcommand{\RS}{\ensuremath{\mathit{RS}}\xspace}

\newcommand{\aTransID}{\ensuremath{\lambda}\xspace}
\newcommand{\TIDIssue}{\ensuremath{\mathsf{issue}()}\xspace}
\newcommand{\TIDRequest}[2]{\ensuremath{\mathsf{request}(#1, #2)}}
\newcommand{\TIDFlush}{\ensuremath{\mathsf{flush}()}\xspace}
\newcommand{\TIDUpdate}[1]{\ensuremath{\mathsf{update}(#1)}}
\newcommand{\TIDRecover}[1]{\ensuremath{\mathsf{recover}(#1)}}
\newcommand{\TIDDrop}[1]{\ensuremath{\mathsf{drop}(#1)}}
\newcommand{\TIDs}[1]{\ensuremath{\Lambda(#1)}}

\newcommand{\TransRelArrow}[1]{\ensuremath{\xrightarrow{~ #1 ~}}}
\newcommand{\TransRel}[3]{\ensuremath{#1 \xrightarrow{~ #2 ~} #3}}

\newcommand{\TRule}[1]{\textsc{T-#1}}
\newcommand{\TRIssue}{\TRule{Iss}\xspace}
\newcommand{\TRRequestStationary}{\TRule{ReqS}\xspace}
\newcommand{\TRRequestTransitioning}{\TRule{ReqT}\xspace}
\newcommand{\TRFlush}{\TRule{Fsh}\xspace}
\newcommand{\TRUpdate}{\TRule{Upd}\xspace}
\newcommand{\TRRecover}{\TRule{Rcv}\xspace}
\newcommand{\TRDrop}{\TRule{Drp}\xspace}

\newcommand{\EffState}[1]{\ensuremath{\mathsf{eff}(#1)}}

\newtheorem{example}{Example}
\newtheorem{lemma}{Lemma}
\newtheorem{definition}{Definition}
\newtheorem{theorem}{Theorem}
\newtheorem{proposition}{Proposition}

\begin{document}


\title{HCAP: A History-Based Capability System for IoT Devices}

\author{Lakshya Tandon \qquad Philip W. L. Fong \qquad Reihaneh
  Safavi-Naini\\
University of Calgary, Alberta, Canada\\
\texttt{\{\,lakshya.tandon,\,pwlfong,\,rei\,\}@ucalgary.ca}
}
\date{March 30, 2018}


\maketitle

\begin{abstract}
  Permissions are highly sensitive in Internet-of-Things (IoT)
  applications, as IoT devices collect our personal data and control
  the safety of our environment.  Rather than simply granting
  permissions, further constraints shall be imposed on permission
  usage so as to realize the Principle of Least Privilege.  Since IoT
  devices are physically embedded, they are often accessed in a
  particular sequence based on their relative physical positions.
  Monitoring if such sequencing constraints are honoured when IoT
  devices are accessed provides a means to fence off malicious
  accesses.  This paper proposes a history-based capability system,
  HCAP, for enforcing permission sequencing constraints in a
  distributed authorization environment.  We formally establish the
  security guarantees of HCAP, and empirically evaluate its
  performance.
\end{abstract}

\section{Introduction}

Internet-of-Things (IoT) devices collect our personal data (e.g.,
wearables, sensors) and control the safety of our environment (e.g.,
thermostat, smart locks).  Granting permissions to access IoT devices
often comes with significant privacy, security and even safety
implications.  Yet, authority delegation is a common use case in IoT
applications. For example, envision the wide deployment of smart locks
on an organization campus.  Permissions to unlock various entrances
must now be properly granted to members of that organization.  In
this work, we are concerned with the potential misuse of permissions
by users of IoT devices.

Rather than simply granting permissions, further usage constraints
shall be imposed on permissions in order to rule out potentially
malicious usage patterns.  Simple examples of this would include
contextual constraints such as not allowing entry after midnight.
Such constraints are a way of realizing the Principle of Least
Privilege \cite{LeastPrivilege}.  But we can do even better than
contextual constraints.  Since IoT devices are physically embedded,
they are often accessed in a particular sequence based on their
relative physical positions.  Monitoring if such sequencing
constraints are honoured provides a means to fence off malicious
accesses to devices.
\begin{example}[Physical Embeddedness] \label{ex-embed} Alice often
  stays after office hour, when all the doors of her organization are
  locked.  Special permissions are granted to her to unlock certain
  doors.  When Alice leaves for the day, she passes through entrances
  $A$, $B$ and $C$ (corresponding respectively to the lab door, the
  computer science building entrance, and the campus gate) in that
  physical order.  Not only does Alice require the permissions to
  unlock the smart locks at $A$, $B$ and $C$, ordering constraints
  shall also be imposed so that, during off hours, $C$ (campus gate)
  is not unlocked before the authorization system registering her
  unlocking $A$ (lab door) and $B$ (building entrance).  Directly
  unlocking $C$ (campus gate) without going through $A$ (lab door) and
  $B$ (building entrance) could very well mean her smartphone has been
  picked up by an unauthorized party trying to enter the campus from
  the outside.
\end{example}

Permission sequencing constraints are also important when device
accesses must conform to a workflow specification.  Kortuem \emph{et
  al.} use the term \Dfn{process awareness} to refer to the ability of
smart objects to guide their users in following operational procedures
\cite{ProcessAwareness}.  Procedure guiding is envisioned to be a key
feature of, say, smart construction objects \cite{SmartConstructionObjects}.
\begin{example}[Process Awareness] \label{ex-process}
  Suppose an industrial process follows an explicitly articulated
  workflow, in which two conceptual steps, $S_1$ and $S_2$, are
  sequentially ordered.  Permissions to operate equipments are
  assigned to each step \cite{Wang-Li:2010, Crampton-etal:2013}: e.g.,
  permissions $p_1$ and $p_2$ can be exercised in step $S_1$, and
  permissions $p_2$ and $p_3$ can be exercised in step $S_2$.  The
  permission assignment and the ordering of workflow steps jointly
  induce sequencing constraints on permission usage: Once $p_3$ is
  exercised, it is obvious that $S_2$ is being executed, and thus
  $p_1$ shall no longer be allowed (i.e., no $p_1$ after $p_3$).
  Violation of this sequencing constraint is a sign of equipment
  misuse.
\end{example}

Constraining the order in which permissions are exercised is in fact
the spirit of History-Based Access Control (HBAC)
\cite{Evans-Twyman:1999,Schneider:2000,IRM,Wallach-etal:2000,Fong:2004,Krukow-etal:2008,Ligatti-etal:2009},
in which the authorization decision of an access request is a function
of the access history.  HBAC policies can be imposed to restrict
permission usage once an access pattern has been detected (no $p_1$
after $p_3$).  This feature can be leveraged for enforcing the
following forms of permission usage control.
\begin{example}[Permission Usage Control]
  Suppose a campus visitor is only allowed to access a facility (e.g.,
  a smart coffee dispenser) no more than four times during her
  visit. In other words, we want the authorization system to deny the
  usage of a permission after it has been exercised for a number of
  times. In short, permissions are seen as consumable resources.  A
  second form of permission usage constraints is the \Dfn{cardinality
    constraint}, which demands that no more than $k$ of the
  permissions in a set $P$ can be exercised \cite{RBAC}.  In the
  special case of $|P| = 2$ and $k = 1$, the constraint enforces
  permission-level \Dfn{mutual exclusion}.  Similarly, the Chinese
  Wall policy \cite{ChineseWall} can be seen as a third form of
  permission usage constraints: Once a resource has been accessed,
  access to resources in conflict with the former will be denied.
\end{example}

Enforcing HBAC policies requires support from the authorization
system.  The growing scale of smart devices and casual users makes it
necessary for device administrators to be able to manage access
control policies centrally, while enabling devices to enforce such
policies in a decentralized manner (i.e., without the mediation of a
centralized reference monitor).  Typically, an unforgeable capability
(aka security token) is issued by a centralized authorization server
to a client, who in turn presents the capability to the resource
custodian as a proof of authorization.  Examples of such distributed
capability systems include the Identity-based Capability System (ICAP)
\cite{ICAP}, CapaFS \cite{Regan-Jensen:2001}, O'Auth \cite{OAuth},
OpenID \cite{OpenID}, and Macaroons \cite{Birgisson-etal:2014}.  The
dual requirements of centralized policy administration and
decentralized policy enforcement are the main driver behind the recent
push by access management solution providers \cite{ForgeRock, Nulli}
to adopt the User-Managed Access standard \cite{UMA} as the choice
platform for access management in IoT applications.  More generally,
distributed capability systems have emerged as a popular choice for
decentralized access control in the IoT literature
\cite{Mahalle-etal:2013, Gusmeroli-etal:2013, Hernandez-Ramos-etal:2016}.
Unfortunately, none of the distributed capability systems surveyed
above offers adequate support for HBAC.  The
crux of the problem is that policy enforcement in HBAC is a stateful
process.  A traditional capability, however, captures a static set of
authorized permissions.  Such capabilities will have to be revoked and
reissued by the centralized authorization server whenever the state of
HBAC enforcement changes, causing the authorization server to be
contacted frequently, thereby nullifying the benefit of
decentralized access control promised by distributed capability
systems.

This paper proposes a History-Based Capability System, HCAP, for
regulating the order in which permissions are exercised in a
distributed authorization environment.  HCAP is an extension of ICAP
\cite{ICAP}.  HCAP capabilities carry sequencing constraints in the
form of security automata (SA) \cite{Schneider:2000}.  Exercising a
permission produces an SA state transition, invalidates the existing
capability, and generates a new capability reflecting the new SA
state.  Since the proposed scheme minimizes communication with the
central authorization server, and does not require the IoT devices to
know about the access control policies, HCAP makes a good building
block for UMA-style combination of centralized policy administration
and decentralized policy enforcement.  We claim four contributions:
\begin{compactenum}[\ (1)]
\item We describe the design of HCAP, a distributed capability system
  that can enforce history-based access control policies
  (see \S \ref{sec-HCAP}).
\item We formally establish the security guarantees of HCAP in the
  form of a safety property and a liveness property (see \S
  \ref{sec-security}).
\item In the formulation of core HCAP, it is assumed that the
  permissions associated with an SA are all related to a single
  device.  We propose an extension to HCAP that relaxes this
  restriction, thereby allowing an SA to regulate the permission usage
  of multiple devices (see \S \ref{sec-multi}).
\item We empirically evaluate the performance of HCAP (see \S
  \ref{sec-performance}).
\end{compactenum}

\section{Related Work}

There are two kinds of distributed authorization systems
\cite{Taly-Shankar:2016}.  The first kind are the
\Dfn{credentials-based authorization systems} (F.~Schneider's
terminology), in which the client presents a set of certificates to an
authorization system as a proof of policy compliance
\cite{Abadi-etal:1993, Appel-Felten:1999, Li-etal:2003,
  Becker-etal:2010}.  The certificates in the compliance proof are
typically issued by different authorities, and each certificate
corresponds to an assertion in a logical language used for specifying
conditions of authorization.  The authorization system is presumed to
know the access control policy, which is specified in the
aforementioned logical language.  The second kind are the
\Dfn{distributed capability systems}, in which the authorization
system, upon successful check of policy compliance, issues to the
client an unforgeable capability (aka security token) \cite{ICAP,
  Regan-Jensen:2001, OAuth, OpenID, Birgisson-etal:2014}. The client
then presents the capability to resource custodians to gain access
without further mediation of the authorization system.  The resource
custodians are thus freed from needing to know and manage the access
control policies.  This work contributes to the literature of the
second kind.  To the best of our knowledge, HCAP is the first
distributed capability system to support History-Based Access Control
(HBAC). We advocate the employment of this feature for sequencing
permission usage in IoT environments.

HCAP is an extension of ICAP \cite{ICAP} in order to enforce HBAC
policies.  While an ICAP capability carries the list of granted
permissions, an HCAP capability carries a partial specification of an
SA, which we call an SA fragment.  In the degenerate case when the SA
has only one state, an HCAP capability is structurally equivalent to
an ICAP capability.  Another point of comparison concerns the
exceptions.  Exceptions are created in ICAP when the authorization
server informs the resource server of capability revocations.  In
HCAP, an exception is created when the resource server exercises a
permission that leads to SA state transition, thereby invalidating an
existing capability.  In addition, HCAP exceptions are much more
complex: each exception chronicles the history of SA state transition,
rather than a single event of revocation.

ICAP has also been extended by Mahalle \emph{et al.}
for IoT applications \cite{Mahalle-etal:2013}. Their extension does
not support HBAC.

The State-Modifying Policy (SMP) language is
an authorization logic that supports the specification of
changes to the protection state as a result of authorization
\cite{Becker-Nanz:2010}.  SMP is designed particularly for 
HBAC policies.  
Its enforcement model, however, assumes a centralized resource guard
(aka PEP), and is therefore incapable of
supporting decentralized access control in the manner of HCAP.

In history-based access control, the history of access is tracked by a
reference monitor, and this access history forms the basis of making
authorization decisions \cite{Evans-Twyman:1999, Schneider:2000, IRM,
  Wallach-etal:2000, Fong:2004, Krukow-etal:2008, Ligatti-etal:2009}.
Schneider proved that only safety properties are enforceable using a
reference monitor that tracks execution history, and proposed the
Security Automata as an automata-theoretic representation of reference
monitors \cite{Schneider:2000}.  In this work, permission sequencing
constraints are encoded as a Security Automaton and embedded in a
capability.  State transition occurs when the capability is presented
to a resource custodian, who may not immediately relay this change
back to the central authorization server.  The representation of the
current automaton state is therefore distributed across multiple
participants.  The technical challenge addressed by HCAP are (a) to
ensure the coherence of this distributed representation, (b) to
provably prevent replay attacks, and (c) to achieve the above while
minimizing communications with the authorization server.

\section{A History-Based Capability System}
\label{sec-HCAP}


\subsection{Overview}
\label{sec-overview}

\paragraph{Protocol Participants.}  We envision a distributed
authorization system akin to UMA \cite{UMA}, consisting of resource
servers, clients and an authorization server.
\begin{asparadesc}
\item[Resource servers.]  Each resource server encapsulates a number
  of resources within a single device, and acts as their custodian.
  For example, a smart weather station tracks a number of weather
  readings, each considered a separate resource.  Operations can be
  performed on the resources.  In the smart weather station, an
  operation for a given weather reading (e.g., temperature) can be
  ``retrieve'' or ``post reading to Facebook.''  A \Dfn{permission} is
  an operation-resource pair.  An access request is a request for the
  resource server to \Dfn{exercise} a permission: i.e., perform the
  operation on the resource.

\item[Clients.]  Clients are users who, mediated by software systems
  (e.g., smartphone apps), direct access requests to
  resource servers.

\item[Authorization server.]  The authorization server is responsible
  for access management.  Administrators of resource servers and their
  resources may specify access control policies to indicate which
  permission is granted to which client.  We assume that an
  Application Programming Interface (API) is in place for an
  administrator to specify resources and operations to be protected.
  What is unique in HCAP is that the access control policy not only
  grants a set of permissions to a client, but also prescribes
  constraints on the order in which the permissions are to be
  exercised.  As we shall see below, such constraints will take the
  form of a \Dfn{security automaton} \cite{Schneider:2000}.
\end{asparadesc}

The authorization server issues capabilities
\cite{Dennis-VanHorn:1983} to clients, who in turn present the
capabilities to resource servers.

\paragraph{Trust Assumptions.} 
The following are assumed.
\begin{inparaenum}[(1)]
\item The authorization server and resource servers are trusted
  parties.
\item Clients are not trusted: they actively attempt to forge, share
  with others, or replay capabilities (and other tickets).  They are
  also unreliable: they may lose capabilities (and other tickets) that
  have been issued to them.
\item A public key infrastructure (PKI) is in place, so that the
  authorization server and the resource servers can authenticate one
  another, as well as the identity claims of clients. It is assumed
  that the authorization server can verify membership of the resource
  servers and clients that belong to the organization.
\item Each resource server has established a shared secret with the
  authorization server.
\item Devices are equipped with a secure untamperable hardware that
  holds their secret values.
\item We assume a central clock that is used to synchronize all
  individual clocks in the system, and assume entities communicate
  over secure channel.
\end{inparaenum}

\paragraph{Design Objectives.}  HCAP is designed with the
following objectives in mind.
\begin{asparadesc}
\item[\Obj{1}] Resource servers shall not maintain knowledge of client
  identities and sequencing constraints.  The authorization server
  alone is responsible for access management.  In other words, when
  access control policies evolve, the resource servers do not need to
  be reconfigured.
\item[\Obj{2}] Communication with the authorization server shall be
  minimized, because that server is a communication bottleneck. A
  protocol in which every access request is mediated by the
  authorization server is considered a non-solution.
\item[\Obj{3}] The computational demand for resource servers shall be
  minimized, as these servers are computationally constrained.
\end{asparadesc}

\paragraph{Solution Approach.} 
The sequencing constraints for permissions are essentially safety
properties \cite{Lamport:1977}, encoded as security automata (an
automata-theoretic representation of reference monitors)
\cite{Schneider:2000}.  When a client initiates a protocol session, a
security automaton is started to monitor the order in which
permissions are exercised in that session.  The authorization server
tracks the current state of that automaton.  Since the resource server
does not know the security automaton (\Obj{1}), and the authorization
server shall not be involved in policy mediation (\Obj{2}), part of or
all of the security automaton is carried in the capability.  When the
client presents the capability to a resource server for gaining
access, the resource server simulates the automaton's state
transition.  State transition is not communicated immediately to the
authorization server (\Obj{2}).  Instead, the resource server records
the state transition, and issues a new capability to the client (while
revoking the previous one).  To conserve the computational resources
of the resource server, the HCAP protocol has the resource server
flushes its record of state transitions back to the authorization
server from time to time, a process known as garbage collection
(\Obj{3}). For the same reason, the capability may not carry the full
specification of the security automaton, thereby making capability
processing lightweight (\Obj{3}).

As the knowledge of the authorization server in the automaton state
may lag behind the state transitions carried out remotely by 
resource servers, the current state of the security automaton is a
datum distributed between the authorization server and the resource
servers.  The design of HCAP ensures the coherence of this distributed
representation of the automaton state, and that outdated capabilities
are not replayed by the client to gain access illegally.

\paragraph{Core HCAP} To facilitate presentation, this and the next
section will present \Dfn{core HCAP}, in which there is only one
resource server. The extension of core HCAP to handle multiple
resource servers is deferred to \S \ref{sec-multi}.

\paragraph{Preliminaries.}
We write $\Dom{f}$ and $\Ran{f}$ respectively for the domain and range
of function $f$.  A \Dfn{partial function} $f : A \PartFunc B$ is a
function $f$ with domain $A' \subseteq A$ and codomain $B$.  A finite
partial function (i.e., finite domain) is also called a \Dfn{map}.

\subsection{Building Blocks}

\paragraph{Security Automata.}

A \Dfn{security automaton} is an automata-theoretic encoding of a
safety property \cite{Schneider:2000}.  Here, we adopt the finitary
variant of deterministic security automaton as defined by Fong
\cite{Fong:2004}.  A \Dfn{Deterministic Finite Security Automaton
  (DFSA)}, or simply \Dfn{Security Automaton (SA)} in this work, is a
tuple $(\SAAlpha, \SAStates, \SAInit, \SATrans)$, where $\Sigma$ is a
finite set of permissions, $Q$ is a finite set of automaton states,
$q_0 \in Q$ is an initial state, and
$\delta: Q \times \Sigma \PartFunc Q$ is a partial transition function
(i.e., $\delta(q, p)$ may be undefined for some pair of state
$q \in Q$ and permission $p \in \Sigma$).  An SA is essentially like a
deterministic finite automaton in which every state is a final state.
An SA accepts a sequence of permissions so long as the transition
function defines a transition for every step.  Policy violation is
detected when there is no transition for a permission in the current
state.  For instance, the two SA below enforce the policies described
in Example \ref{ex-embed} (left) and Example \ref{ex-process} (right).
\[
\xymatrix{
  \ar@{~>}[r] 
  &
  q_0 
  \ar[r]^{A}
  & 
  q_1
  \ar[r]^{B} 
  & 
  q_2 
  \ar[r]^{C} 
  & 
  q_3
}
\qquad\qquad
\xymatrix@C=1.4pc{
 \ar@{~>}[r]
 &
 q_0 
 \ar[r]^{p_3}
 \ar@(ul,ur)^{p_1, p_2}
 & 
 q_1
 \ar@(ul,ur)^{p_2, p_3}
}
\]

When the SA \anSA is in a state \anSAState, the permissions that bring
\anSA back to \anSAState are \Dfn{stationary} permissions, and the
permissions that cause a transition to a different state are
\Dfn{transitioning} permissions.  We define
$\SAStationary{ \anSA }{ \anSAState } = \{ \aPerm \in \SAAlpha \mid
\SATrans( \anSAState, \aPerm ) = \anSAState \}$ to be the set of
stationary permissions in state \anSAState.  Similarly, we define
$\SATransitioning{ \anSA }{ \anSAState } = \{ \aPerm \in \SAAlpha \mid
\exists \anSAState' \in \SAStates \,.\, \anSAState' \neq \anSAState
\land \SATrans( \anSAState, \aPerm ) = \anSAState' \}$ to be the set
of transitioning permissions in state \anSAState.

\paragraph{Tickets.}  The authorization server and resource server
issue \Dfn{tickets} to the client, who in turn uses them to justify
requests.  To ensure ticket authenticity and non-transferability,
each ticket carries an authentication tag obtained from the shared
secret \aKey and the client identity \aUID.  In this work, we follow
the lightweight tagging mechanism of ICAP \cite{ICAP}. Suppose
\anAssertion is an assertion, the ticket \Encap{ \aKey }{ \aUID }{
  \anAssertion } signed by secret key \aKey for client \aUID is
$\anAssertion \mid h( \anAssertion \mid \aUID \mid \aKey )$, where $h$
is a hash function and ``$x \mid y$'' means the concatenation of $x$
and $y$.  Upon receiving a ticket \Encap{ ? }{ ? }{ \anAssertion }
from a client \aUID, one can check its authenticity by checking that
the hash value in \Encap{ ? }{ ? }{ \anAssertion } is equal to
$h( \anAssertion \mid \aUID \mid \aKey )$.  The use of \aUID to
compute the hash value also ensures that the ticket is
non-transferrable.

There are two kinds of tickets, capabilities
and update requests, which we introduce in turn.

\paragraph{Capabilities.}
Capabilities are tickets issued by either the authorization server or
the resource server to assert that certain permissions can be
exercised by the client.  A client may present an access request along
with the capability to gain access.  A capability issued to client
\aUID for session \aSessID has the form \EncapCap{ \aKey }{ \aUID }{
  \aSessID }{ \TSSerial }{ \anSAFrag }.  Every capability
asserts that the SA is in a certain state \anSAState, and thus some
corresponding permissions can be exercised for \aSessID.  The
timestamp \TSSerial, also called the \Dfn{serial number} of the
capability, identifies \anSAState indirectly by identifying the time
when the SA entered into state \anSAState.  
The component \anSAFrag is an
\Dfn{SA fragment}, which identifies the permissions (stationary and
transitioning) allowed in state \anSAState, as well as the transitions
emanating from \anSAState.  In some sense, an SA fragment is a partial
specification of the SA, with current state \anSAState (formal
definition to be given below).  The specification is partial, because
the authorization server is not obligated to encode the entire SA in
one capability.  This may be because full encoding causes the
capability to be bloated, or because the authorization server desires
to be synchronized with the resource server more often, or because the
underlying communication protocol limits the capability size.

\paragraph{SA Fragments.}
An SA fragment is a representation of two things: (a) a (possibly
incomplete) transition diagram, and (b) the current state of the
transition diagram. States in the transition diagram are identified by
symbolic names. We assume there is a countably infinite set \Names of
symbolic names as well as a distinct marker \Unknown (pronounced
`unknown') such that $\Unknown \not\in \Names$.  An SA fragment
\anSAFrag is a pair \SAFrag{ \FDefs }{ \aCurrName }.  The component
\FDefs is a finite partial function in which
$\Dom{ \FDefs } \subset \Names$ identifies the states of a transition
diagram. For each name $\aName \in \Dom{ \FDefs }$,
$\FDefs ( \aName )$ specifies the transitions emanating from the state
with name \aName.  More specifically, $\FDefs ( \aName )$ is a pair
$( \SP, \FTrans )$, so that $\SP \subseteq \SAAlpha$ is the set of
stationary permissions of \aName, and
$\FTrans : \SAAlpha \PartFunc \Names \cup \{ \Unknown \}$ maps each
transitioning permission of \aName to either a next state or the
marker \Unknown. When $\FTrans ( \aPerm ) = \Unknown$, the transition
diagram permits the transition but does not identify the next state of
the transition (thus \anSAFrag is a fragment rather than a complete
SA).  We further require that $\SP \cap \Dom{ \FTrans } = \emptyset$,
and
$\Ran{ \FTrans } \setminus \{ \Unknown \} \subseteq \Dom{ \FDefs }$.
Lastly, the second component \aCurrName of \anSAFrag identifies the
current state of the transition diagram, such that
$\aCurrName \in \Dom{ \FDefs }$.  It is easy to see that one can use
an SA fragment to partially specify an SA (i.e., a subset of states
plus a subset of transitions). An SA fragment can be encoded as a JSON
(JavaScript Object Notation) object in a straightforward manner
\cite{JSON}.

Transitions can be computed efficiently when SA fragments are encoded
in JSON.  Given an SA fragment
$\anSAFrag = \SAFrag{ \FDefs }{ \aCurrName }$ for which
$\FDefs ( \aCurrName ) = ( \SP, \FTrans )$, \FDo{ \anSAFrag }{ \aPerm
} is defined to be (a) \anSAFrag if $\aPerm \in \SP$, (b) the fragment
\SAFrag{ \FDefs }{ \FTrans(\aPerm) } if $\FTrans(\aPerm) \in \Names$,
and (c) \Unknown if $\FTrans(\aPerm) = \Unknown$.  Otherwise \FDo{
  \anSAFrag }{ \aPerm } is undefined.

Our security guarantees depend on the condition that the SA fragments
embedded in capabilities are ``conservative'' partial specification of
the corresponding SA: i.e., the SA fragment does not allow transitions
that are not supported by the corresponding SA, a notion that we
formalize in the following.  Let
$\anSAFrag = \SAFrag{ \FDefs }{ \aCurrName }$ be an SA fragment,
$\anSA = (\SAAlpha, \SAStates, \SAInit, \SATrans)$ be an SA, and
$\anSAState \in \SAStates$ be an SA state.  Then \anSAFrag is
\Dfn{safe for \anSA in state \anSAState} if and only if there exists a
function $\pi : \Dom{ \FDefs } \rightarrow \SAStates$ such that (a)
$\pi ( \aCurrName ) = \anSAState$, and (b) for every
$\aName \in \Dom{ \FDefs }$, where
$\FDefs ( \aName ) = ( \SP, \FTrans )$, the three conditions below
hold: (i) $\SP \subseteq \SAStationary{ \anSA }{ \pi( \aName ) }$;
(ii)
$\Dom{ \FTrans } \subseteq \SATransitioning{ \anSA }{ \pi( \aName )
}$; (iii) for every $\aPerm \in \Dom{ \FTrans }$, either
$\FTrans( \aPerm ) = \Unknown$ or
$\pi( \FTrans ( \aPerm ) ) = \SATrans ( \pi( \aName ), \aPerm)$.

\begin{lemma} \label{lemma-fragment} Suppose $\anSA = (\SAAlpha,
  \SAStates, \SAInit, \SATrans)$ is an SA, $\anSAState
  \in \SAStates$, and SA fragment $\anSAFrag = \SAFrag{ \FDefs }{
    \aCurrName }$ is safe for \anSA in \anSAState. Let 
   $\FDefs ( \aCurrName ) = ( \SP, \FTrans )$. Then the following
  properties hold:
\begin{inparaenum}
\item $\FDo{ \anSAFrag }{ \aPerm }$ is defined only if
  $\SATrans (\anSAState, \aPerm ) $ is defined.
\item If $\aPerm \in \SP$, then \aPerm is stationary
  for \anSAState. If $\aPerm \in \Dom{ \FTrans }$, then
  \aPerm is transitioning for \anSAState.
\item If $\FDo{ \anSAFrag }{ \aPerm }$ is an SA fragment (rather than
  \Unknown), then \FDo{ \anSAFrag }{ \aPerm } is safe for \anSA in
  $\SATrans ( \anSAState, \aPerm )$.
\end{inparaenum}
\end{lemma}

\paragraph{Update Requests.}  A second kind of tickets is an
\Dfn{update request}, which has the form \EncapUpdate{ \aKey }{ \aUID
}{ \aSessID }{ \anEx }. An update request is issued by the resource
server, asserting that since last synchronized with the authorization
server, \anEx is the list of transitioning permissions that have been
exercised by the resource server for the session \aSessID.  The
construct \anEx is called an \Dfn{exception}, for it describes how the
knowledge of the authorization server has been out of sync.

\paragraph{Exceptions.}
An \Dfn{exception} \anEx records the history of the resource server
having exercised certain permissions in the past.  It is defined
inductively as follows:
\[
   \anEx ::= \ExNil{ \aTS } \mid \ExEx{ \aPerm }{ \aTS }{ \anEx }
\]
where $\aPerm \in \SAAlpha$ and $\aTS \in \Nat$.  Essentially, \anEx
is a list of permission-timestamp pairs.  Each pair contains a
permission \aPerm and the time \aTS at which \aPerm was exercised.
The permissions are listed in descending order of time (more recent
ones are listed first).  We write \Times{ \anEx } for the set of all
timestamps appearing in \anEx, as well as \TSFirst{ \anEx } and
\TSLast{ \anEx } respectively for the minimum (least recent) and
maximum (most recent) timestamps in \Times{ \anEx }.

We write $\SATransEx ( \anSAState, \anEx)$ to signify the SA state
obtained by starting at state \anSAState and exercising the
permissions of \anEx in chronological order.  That is, $\SATransEx (
\anSAState, \ExNil{ \aTS }) = \anSAState$, and $\SATransEx (
\anSAState, \ExEx{ \aPerm }{ \aTS }{ \anEx } ) = \SATrans (\SATransEx
( \anSAState, \anEx ), \aPerm)$.  \SATransEx is undefined if one of
the recursive calls is undefined.

A similar notation, \FDoEx{ \anSAFrag }{ \anEx }, can also be defined
for SA fragments: $\FDoEx{ \anSAFrag }{ \ExNil{ \aTS } } = \anSAFrag$,
and $\FDoEx{ \anSAFrag }{ \ExEx{ \aPerm }{ \aTS }{ \anEx } } = \FDo{
  \FDoEx{ \anSAFrag }{ \anEx } }{ \aPerm }$. As expected, \FDoEx{
  \anSAFrag }{ \anEx } is not defined if the nested calls are not
defined, or if they return \Unknown.

Sometimes we want to apply only some of the transitions in an
exception list to an SA fragment.  Suppose $\anEx = \ExEx{ \aPerm_m }{
  \aTS_m }{ \ExEx{ \aPerm_{m-1} }{ \aTS_{m-1} }{ \ldots
    \ExEx{ \aPerm_1 }{ \aTS_1 }{ \ExNil{ \aTS_0 }} \ldots } }$ for
some $m \geq 0$.  We then write \FDoExUpto{ \aTS_i }{ \anSAFrag }{
  \anEx } to denote \FDo{ \ldots \FDo{ \FDo{ \anSAFrag }{ \aPerm_1 }
  }{ \aPerm_2 } \ldots }{ \aPerm_i }, and \FDoExAfter{ \aTS_i }{
  \anSAFrag }{ \anEx } to denote \FDo{ \ldots \FDo{ \FDo{ \anSAFrag }{
      \aPerm_{i+1} } }{ \aPerm_{i+2} } \ldots}{ \aPerm_m }. It is not
hard to see that $\FDoEx{ \anSAFrag }{ \anEx } = \FDoExAfter{ \aTS_i
}{ \FDoExUpto{ \aTS_i }{ \anSAFrag }{ \anEx } }{ \anEx }$.

\subsection{Protocol Description}

\paragraph{Server Internal States.}  
The authorization server and the resource server maintain a shared
secret \aKey.  In addition, the authorization server maintains three
maps for session administration:  (a) \MapMonitor{ \aSessID
} is the SA for session \aSessID, (b) \MapState{ \aSessID } is the
state of \MapMonitor{ \aSessID } last known by the authorization
server, and (c) \MapSerial{ \aSessID } is the timestamp when
\MapMonitor{ \aSessID } is registered by the authorization server to
have entered into state \MapState{ \aSessID }.  The timestamp
\MapSerial{ \aSessID } will be used as the serial number of the next
capability issued by the authorization server for session \aSessID.
Lastly, the authorization server precomputes an SA fragment
\MapFragment{ \anSA }{ \anSAState } for every SA \anSA stored in
\MapMonitor{ \cdot } and every state \anSAState of \anSA. It is
assumed that \MapFragment{ \anSA }{ \anSAState } is safe for \anSA in
\anSAState.

The resource server maintains two pieces of information: (a) a
timestamp \TSValid, which marks the minimum serial number of
capabilities that the server considers valid, and (b) a map \MapEx{
  \cdot }, which records, for each known session ID \aSessID, the
exception that chronicles the transitioning permissions the resource
server has exercised since the SA of \aSessID has entered the state
\MapState{ \aSessID }.

\paragraph{Session Initialization.}
A client \aUID who intends to access a resource server shall first
authenticate itself to the authorization server, and then request the
initiation of a new protocol session for that resource server.  The
authorization server will consult an access control policy, and decide
if access shall be granted.\footnote{The authorization decision can
  take into account identities, roles (RBAC) \cite{RBAC}, attributes
  (ABAC) \cite{ABAC} and relationships (ReBAC) \cite{Fong:2011}.} If
the authorization decision is positive, a new session ID \aSessID is
created.  The access control policy will grant a set \SAAlpha of
permissions to the session, and also prescribe an SA
$\anSA = \SA{ \SAAlpha }{ \SAStates }{ \SAInit }{ \SATrans }$ to
regulate the order in which permissions are to be exercised within
that session.  The session \aSessID is initialized as follows:
\begin{tabbing}
xxx\=\kill
  \>$\MapMonitor{ \aSessID } \leftarrow \anSA$; \\
  \>$\MapState{ \aSessID }  \leftarrow \SAInit$; \\
  \>$\MapSerial{ \aSessID }  \leftarrow \mathit{current\_time}()$;
\end{tabbing}
The capability \EncapCap{ \aKey }{ \aUID }{ \aSessID }{ \TSSerial
}{ \anSAFrag } is then issued to client \aUID, where \aKey is
the shared secret between the authorization and the resource server,
$\TSSerial = \MapSerial{ \aSessID }$, and $\anSAFrag = \MapFragment{
  \anSA }{ \SAInit }$.

\begin{algorithm}
\KwIn{A client access request $(\aUID, \aPerm, \aCap%
)$, where
   \aUID is the client's authenticated identity,
   \aPerm is the permission to be exercised, and
   \aCap is a capability \EncapCap{ ? }{ ? }{ \aSessID 
    }{ \TSSerial }{ \anSAFrag }
    for session
    \aSessID, 
    such that $\anSAFrag = \SAFrag{ \FDefs }{ \aCurrName }$ and
    $\FDefs ( \aCurrName ) = ( \SP, \FTrans )$.
}
\KwOut{A set of tickets, or a
  failure response.}
\KwData{The resource server maintains the following persistent data:
  (a) a secret \aKey it shares with the authorization server,
  (b) a timestamp \TSValid, 
  and
  (c) a map \MapEx{\cdot} that assigns an exception to each
  known
   session ID.
}
 \If{\label{line-integrity}\aCap is not signed by \aKey for \aUID,
       or $\TSSerial < \TSValid$
 }{
  \Return{failure}
 }
 \If{\label{line-ex-first}\label{line-overwrite-ex}\MapEx{ \aSessID } is
   undefined, or $\TSSerial > \TSLast{ \MapEx{ \aSessID } }$}{%
   $\MapEx{ \aSessID } \leftarrow \ExNil{ \TSSerial }$%
 }
 \ElseIf{\label{line-too-old}$\TSSerial < \TSLast{ \MapEx{ \aSessID } }$}{%
     \Return{failure}\label{line-ex-last}%
 }

    \If{\label{line-transition-begin}\label{line-stationary-first}$\aPerm \in \SP$}{
      Exercise permission \aPerm\;
      \Return{\label{line-stationary-last}$\emptyset$}
    }
    \ElseIf{\label{line-transitioning-first}$\aPerm \in \Dom{ \FTrans }$}{
       Exercise permission \aPerm\;
       $\aTS \leftarrow \mathit{current\_time}()$\;
       $\MapEx{ \aSessID } \leftarrow 
              \ExEx{\aPerm}{\aTS}{\MapEx{ \aSessID }}$\label{line-ex-update}\;
       $\anSAFrag' \leftarrow \FDo{ \anSAFrag }{ \aPerm }$\label{line-frag-trans}\;
       \If{$\anSAFrag' = \Unknown$}{
         \Return{\label{line-issue-upd}$\{ 
          \EncapUpdate{ \aKey }{\aUID}{\aSessID}{\MapEx{\aSessID}} \}$}
       }
       \Else{
         \Return{\label{line-transitioning-last}\label{line-issue-cap}$\{ \EncapCap{ \aKey }{\aUID}{\aSessID}{\aTS}{\anSAFrag'} \}$}
       }
    } 
    \lElse{%
       \Return{failure}\label{line-deny}\label{line-transition-end}
    }
\caption{\label{algo-auth}Authorization procedure of the resource server.}
\end{algorithm}

\paragraph{Authorization.}
The client \aUID requests the resource server to exercise a permission
\aPerm by presenting a triple $(\aUID, \aPerm, \EncapCap{ ? }{ ? }{
  \aSessID }{ \TSSerial }{ \anSAFrag })$, where the capability is the
justification for access.  The identity of \aUID is first
authenticated, and then the request is authorized according to
Algorithm \ref{algo-auth}.  The resource server first checks the
authenticity of the capability (line \ref{line-integrity}). It also
rejects capabilities with serial numbers earlier than \TSValid. As we
shall see below, \TSValid is the time of the last garbage collection,
whereby the authorization server and the resource server synchronize
their knowledge of the SA's current states.  Such a synchronization
invalidates all capabilities with serial numbers earlier than
\TSValid.

The serial number of the capability is then compared to 
\TSLast{ \MapEx{ \aSessID } } in lines
\ref{line-ex-first}--\ref{line-ex-last}.  The goal is to see how
current the capability is in comparison to the knowledge of the
resource server.  Line \ref{line-overwrite-ex} corresponds to the case
when the capability is issued by the authorization server, and the
latter's knowledge of the SA state is more current than that of the
resource server.  Consequently, \MapEx{ \aSessID } is reset.  Line
\ref{line-too-old} corresponds to the case when the capability
captures an SA state that is older than what the resource server
knows. The capability is therefore rejected.  The fall-through case of
lines \ref{line-ex-first}--\ref{line-ex-last} is when \TSSerial equals
\TSLast{ \MapEx{ \aSessID } }, meaning that the capability is as
current as the knowledge of the resource server, and nothing needs to
be done in this case.

Lines \ref{line-stationary-first}--\ref{line-stationary-last} handle
requests that involve the exercising of stationary permissions.  No
new ticket is issued.

Lines \ref{line-transitioning-first}--\ref{line-transitioning-last}
specify the case when the request involves a transitioning permission.
The permission is exercised, and the transition is recorded in \MapEx{
  \aSessID } (line \ref{line-ex-update}).  The provided SA fragment is
then used for computing the next SA fragment (line
\ref{line-frag-trans}).  A new capability is issued for the new SA
fragment (line \ref{line-issue-cap}).  Line \ref{line-issue-upd} will
be discussed below under the heading \emph{Update Requests}.

If the permission requested is neither stationary nor transitioning,
then the request is denied (line \ref{line-deny}).

\paragraph{Update Requests.}
An update request is issued when the SA fragment embedded in the
capability does not provide enough information for the resource server
to construct the next capability (line \ref{line-issue-upd}).  The
client is expected to take the update request to the authorization
server, so that the latter can update its record of the current SA
state.  The authorization server will only accept an update request
\EncapUpdate{ \aKey }{ \aUID }{ \aSessID }{ \anEx } if $\TSFirst{
  \anEx } = \MapSerial{ \aSessID }$.  The result is that \MapSerial{
  \aSessID } is updated to the current time, and \MapState{ \aSessID }
is updated to $\SATransEx( \MapState{ \aSessID }, \anEx )$, where
\SATrans is the transition function of \MapMonitor{ \aSessID }.
Lastly, the authorization server will issue to the client a fresh
capability, in which the SA fragment is \MapFragment{ \MapMonitor{
    \aSessID } }{ \MapState{ \aSessID } }.

\paragraph{Garbage Collection.}
The resource server accumulates exception information in \MapEx{ \cdot
} due to the creation of new sessions and exercising transitioning
permissions.  Tracking exception information strains the resource
server, which is hosted on constrained hardware.  That is why
``garbage collection'' needs to be performed from time to time. This
involves the resource server (a) sending the contents of \MapEx{ \cdot
} (e.g., encoded as a JSON object) to the authorization server, (b)
resetting \MapEx{ \cdot } to an empty map, and (c) setting \TSValid to
the current time (i.e., time of garbage collection).  Step (c)
invalidates the tickets issued prior to garbage collection, forcing
clients to obtain fresh capabilities from the authorization server.

Upon receiving \MapEx{ \cdot }, the authorization server updates the
SA state on record for each \aSessID defined in \MapEx{ \cdot }.  More
specifically, \MapState{ \aSessID } is updated to $\SATransEx (
\MapState{ \aSessID }, \linebreak \MapEx{ \aSessID } )$, and 
\MapSerial{ \aSessID } to the time of garbage collection.

\paragraph{Ticket Recovery.} Clients are unreliable, and may
accidentally misplace tickets.  As tickets
contain crucial information for the proper execution
of the protocol, two mechanisms of ticket recovery are in place.

First, a client may contact the authorization server, and have the
capability with serial number \MapSerial{ \aSessID } reissued.  The SA
fragment of that capability is \linebreak \MapFragment{ \MapMonitor{ \aSessID }
}{ \MapState{ \aSessID } }.  This is the standard means for obtaining
a working capability after garbage collection.

The method above may not be sufficient for recovering the latest
tickets.  In particular, if transitioning requests have already been
made since \MapSerial{ \aSessID }, then the resource server has issued
more recent tickets.  So a second ticket recovery mechanism is in
place, in which the client may present to the resource server a
previously issued capability \EncapCap{ \aKey }{ \aUID }{ \aSessID }{
  \TSSerial }{ \anSAFrag }, such that $\TSSerial \in \Times{ \MapEx{
    \aSessID } }$, and then the resource server will use \MapEx{
  \aSessID } together with \anSAFrag to reconstruct the latest ticket
for \aSessID (either an update request or a capability).  Details of
this mechanism are provided in the next section in transition rule
\TRRecover.  The latest ticket for a session can always be recovered
by applying the first and second recovery mechanism in sequence.

\subsection{Discussions}

\paragraph{Authorization server.}
The authorization server has freedom to construct any SA fragment as
\MapFragment{ \anSA }{ \anSAState } so long as the fragment
is safe for \anSA in \anSAState.  The following are some
possibilities:
\begin{compactitem}[~$\bullet$\ ]
\item If the SA \anSA contains a single state, then the HCAP protocol
  degenerates to ICAP, as all permissions are stationary.
\item Consider the case when every capability issued by the
  authorization contains an SA fragment $\anSAFrag = \SAFrag{ \FDefs
  }{ \aCurrName }$ such that $\FDefs ( \aCurrName ) = ( \SP, \FTrans
  )$, $\Dom{ \FDefs } = \{ \aCurrName \}$, and $\Ran{ \FTrans } = \{ \Unknown \}$.
  That is, all transition targets are
  unknown.
  Such a capability
  describes only the stationary and transitioning permissions (via \SP
  and \Dom{ \FTrans } respectively) of the current SA state.  This
  results in highly lightweight capabilities, and the processing
  overhead for the resource server is minimized.  An update request
  will be returned every time a transitioning permission is exercised,
  thereby forcing the client to communicate with the authorization
  server whenever a transition occurs.
\item Capabilities may be constructed to capture several
  levels of transition. For example, if it is known that the most
  frequent transitions will oscillate among a small number of states,
  then that region of the transition diagram can be embedded in the
  capability.  Update requests will be returned infrequently.
\item If \anSA is small, then the entire specification can be
  captured in the capability. No update requests will ever be returned.
\end{compactitem}

\paragraph{Resource server.}
There are two approaches to decide when garbage collection should be
triggered:
\begin{compactenum}[(1)]
\item Garbage collection can be invoked on regular intervals (e.g.,
  every 8 hours). This ensures that sessions that are no longer active
  will not occupy resources indefinitely.
\item Garbage collection can also be invoked when \MapEx{ \cdot }
  reaches a certain size threshold, or when one of the exception lists
  exceeds a certain length threshold. With this approach, \MapEx{
    \cdot } is guaranteed to never grow beyond a predetermined
  capacity.
\end{compactenum}
A combination of both approaches is recommended for a realistic
implementation: perform garbage collection in regular intervals as
well as when the capacity/length threshold is reached.

\section{Security Guarantees}
\label{sec-security}

Replay attacks are the main security concern for HCAP: Is it possible
for the client to gain illegal access by presenting a previously
issued capability to a resource server after the SA has already
transitioned to a state \anSAState in which that capability is no
longer representative of \anSAState.  In this section, we formulate a
formal model for an HCAP protocol session, and demonstrate that replay
attacks are impossible (Safety). We also demonstrate that the protocol
is resilient to unreliable clients who misplace tickets (Liveness).

We model the HCAP protocol as a state transition system.  Each
protocol state captures the state of the entire distributed
authorization system, including the internal states of the
authorization server, the resource server and the client.  A protocol
state transition occurs when the protocol participants interact with
one another.  The main goal of verification is to establish a
correspondence between the distributed authorization system and a
reference monitor that runs in a centralized system (Theorem
\ref{thm-safety}).

Our state transition model abstracts away the following aspects of HCAP:
\begin{inparaenum}[(1)]
\item Ticket forging is not modelled as we assume that it is
   adequately prevented by authentication tags.
\item As protocol sessions are independent from one another, the
  model specifies the behaviour of one protocol session only.
\item We omit the minor detail of \MapEx{ \aSessID } becoming
  undefined after garbage collection.
\end{inparaenum}

\paragraph{Protocol States.}
Throughout this section, we assume that $\anSA = \SA{ \SAAlpha }{
  \SAStates }{ \SAInit }{ \SATrans }$ is the SA for the protocol
session being modelled. 
We further assume that the authorization server
has pre-computed, for each state $\anSAState \in \SAStates$, 
an SA fragment $\anSAFrag_{\anSA,  \anSAState}$
that is safe for \anSA in \anSAState.
\begin{definition}[Protocol States]
  A \Dfn{protocol state} \aPState is a 4-tuple \PState{ \TSClock }{
    \anAState }{ \anRState }{ \aCState }, where the four components
  are defined as follows.
\begin{compactitem}[\ $\bullet$\ ]
\item The component $\TSClock \in \Nat$ is the \Dfn{global clock value}.

\item The \Dfn{authorization server state} \anAState is a pair
  \AState{ \anSAStateA }{ \TSState}, where $\anSAStateA \in \SAStates$
  is the state of \anSA last known by the authorization server,
  and $\TSState \in \Nat$ is the time when the above knowledge is
  registered by the authorization server.
  
\item The \Dfn{resource server state} \anRState is a pair \RState{
    \TSValid }{ \anExR }, where $\TSValid \in \Nat$ is the minimum
  serial number for capabilities that the resource server considers
  valid, and exception \anExR records the transitioning permissions
  that have been exercised by the resource server since \anSA enters
  into state \anSAStateA.

\item The \Dfn{client state} \aCState is the set of \Dfn{tickets}
  that have been issued to the client throughout the protocol session.
  A ticket \aTicket is of one of two forms: (a) an update request
  \CUpdate{ \anEx }, where \anEx is of the form \ExEx{ \aPerm }{ \aTS
  }{ \anEx' }, or (b) a capability \CCap{ \TSSerial }{ \anSAFrag }.

\end{compactitem}
Let \PStates{ \anSA } be the set of all protocol states \aPState of the above form.
\end{definition}

\paragraph{Initial State.}  The protocol is intended to begin at the
initial state
$\aPState_0 = \PState{ 2 }{ \anAState_0 }{ \anRState_0 }{ \emptyset }$
where $\anAState_0 = \AState{ \SAInit }{ 1 }$, $\anRState_0 = \RState{ 1 }{
  \ExNil{ 0 } }$.

\paragraph{State Transition.}  A \Dfn{transition identifier} \aTransID
identifies a protocol event that causes a change to the protocol
state:
\[
\aTransID  ::= 
\TIDIssue \mid
\TIDRequest{ \aPerm }{ \aTicket } \mid
\TIDFlush \mid
\TIDUpdate{ \aTicket } \mid 
  \TIDRecover{ \aTicket } \mid
\TIDDrop{ \aTicketSet }
\]
where \aPerm is a permission, \aTicket is a ticket, and \aTicketSet
is a set of tickets.  Let \TIDs{ \anSA } be the set of all
transition identifiers induced by \anSA.

We specify below a transition relation
$\TransRel{ \cdot }{ \cdot }{ \cdot } \subseteq \PStates{ \anSA }
\times \TIDs{ \anSA } \times \PStates{ \anSA }$.  The relation is
specified in terms of transition rules, which identify the
conditions under which \TransRel{ \PState{ \TSClock }{ \anAState }{
    \anRState }{ \aCState } }{ \aTransID }{ \PState{ \TSClock' }{
    \anAState' }{ \anRState' }{  \aCState' } }, where
$\anAState = \AState{ \anSAStateA }{ \TSState }$,
$\anRState = \RState{ \TSValid }{ \anExR }$,
$\anAState' = \AState{ \anSAStateA' }{ \TSState' }$, and
$\anRState' = \RState{ \TSValid' }{ \anExR' }$.  By default,
$\TSClock' = \TSClock +1$, $\anAState' = \anAState$,
$\anRState' = \anRState$ and $\aCState' = \aCState$, unless the rules
explicitly say otherwise.
\begin{description}

\item[\TRIssue] \emph{The authorization server issues a capability to
    the client.}
 \begin{description}
   \item[Precondition:] $\aTransID = \TIDIssue$
   \item[Effect:] $\aCState' = \aCState \cup \{\, 
     \CCap{ \TSState }{ \anSAFrag_{\anSA, \anSAStateA} } \,\}$.
 \end{description}

\item[\TRRequestStationary] \emph{The client requests
     to exercise a stationary permission.}
 \begin{description}
  \item[Precondition:] $\aTransID = \TIDRequest{ \aPerm }{ \aTicket }$,
    $\aTicket \in \aCState$,
    $\aTicket = \CCap{ \TSSerial }{ \anSAFrag }$, 
    $\TSSerial \geq \TSValid$,
    $\TSSerial \geq \TSLast{ \anExR }$,
    $\anSAFrag = \SAFrag{ \FDefs }{ \aCurrName }$, 
    $\FDefs ( \aCurrName ) = (\SP, \_)$,
    $\aPerm \in \SP$.
  \item[Effect:] 
    $\anExR' = \ExNil{ \TSSerial }$ if $\TSSerial > \TSLast{ \anExR }$.
 \end{description}

\item[\TRRequestTransitioning] \emph{The client requests to
    exercise a transitioning permission.}
 \begin{description}
  \item[Precondition:] $\aTransID = \TIDRequest{ \aPerm }{ \aTicket }$,
    $\aTicket \in \aCState$,
    $\aTicket = \CCap{ \TSSerial }{ \anSAFrag }$, 
    $\TSSerial \geq \TSValid$,
    $\TSSerial \geq \TSLast{ \anExR }$,
    $\anSAFrag = \SAFrag{ \FDefs }{ \aCurrName }$, 
    $\FDefs ( \aCurrName ) = (\_, \FTrans)$,
    $\aPerm \in \Dom{ \FTrans }$.
  \item[Effect:] 
    First, $\anExR' = \ExEx{ \aPerm }{ \TSClock }{ \anEx_0 }$,
    where $\anEx_0 = \ExNil{ \TSSerial }$ if $\TSSerial > \TSLast{
      \anExR }$, or $\anEx_0 = \anExR$ if $\TSSerial = \TSLast{ \anExR
    }$.
    Second, $\aCState' = \aCState \cup \{\, \aTicket_0 
    \,\}$, where 
     $\aTicket_0 = \CUpdate{ \anExR' }$ if $\FDo{
     \anSAFrag }{ \aPerm } = \Unknown$, or
     $\aTicket_0 = \CCap{ \TSClock }{ \FDo{
     \anSAFrag }{ \aPerm } }$ otherwise.
 \end{description}

\item[\TRFlush] \emph{Garbage collection.}
 \begin{description}
  \item[Precondition:] $\aTransID = \TIDFlush$.
  \item[Effect:] First, $\TSState' = \TSClock$,
    $\TSValid' = \TSClock$, and $\anExR' = \ExNil{ \TSLast{ \anExR } }$.
    Second, $\anSAStateA' = \SATransEx( \anSAStateA, \anExR )$ if 
   $\TSState = \TSFirst{ \anExR }$.
 \end{description}

\item[\TRUpdate] \emph{The client updates the internal state of the
    authorization server.}
 \begin{description}
  \item[Precondition:] $\aTransID = \TIDUpdate{ \aTicket }$,
    $\aTicket \in \aCState$, $\aTicket = \CUpdate{ \anEx }$, and
    $\TSFirst{ \anEx } = \TSState$. 
  \item[Effect:]
    $\TSState' =  \TSClock$,
    $\anSAStateA' = \SATransEx( \anSAStateA, \anEx )$.
 \end{description}

\item[\TRRecover] \emph{The client asks the resource server to recover
    a lost ticket.}
 \begin{description}
  \item[Precondition:]
   $\aTransID = \TIDRecover{ \aTicket }$, $\aTicket \in \aCState$,
   $\aTicket = \CCap{ \TSSerial }{ \anSAFrag }$, and $\TSSerial \in \Times{
     \anExR }$.
  \item[Effect:]
   Let $\anSAFrag' = \FDoExAfter{ \TSSerial }{ \anSAFrag }{ \anExR }$.
   There are three cases.
   (1) If $\anSAFrag'$ is undefined, then $\aCState' = \aCState$.
   (2) If $\anSAFrag' = \Unknown$,
   then $\aCState' = \aCState \cup \{ 
    \CUpdate{ \anExR }
   \}$.
   (3) Otherwise, $\aCState' = \aCState \cup \{\, 
    \CCap{ \TSLast{ \anExR } }{ \anSAFrag' }
   \,\}$.
 \end{description}
\item[\TRDrop] \emph{The client accidentally drops some of its tickets.}
 \begin{description}
  \item[Precondition:] $\aTransID = \TIDDrop{ \aTicketSet }$, $\aTicketSet
    \subseteq \aCState$.
  \item[Effect:] $\aCState' = \aCState \setminus \aTicketSet$.
 \end{description}

\end{description}

Appendix \ref{app-invariants} enumerates the state invariants that are
satisfied by the initial state and perserved by the state transition relation.

\paragraph{Security Properties.}
Consider protocol state
$\aPState = \PState{ \TSClock }{ \anAState }{ \anRState }{ \aCState
}$, where $\anAState = \AState{ \anSAStateA }{ \TSState }$ and
$\anRState = \RState{ \TSValid }{ \anExR }$, such that \aPState
satisfies the state invariants in Appendix \ref{app-invariants}.  The
\Dfn{effective SA state} of protocol state \aPState, denoted
\EffState{ \aPState }, is \anSAStateA if
$\TSState > \TSLast{ \anExR }$, or
$\SATransEx ( \anSAStateA, \anExR )$ otherwise.  The internal states
of the authorization and resource server is a distributed
representation of the effective SA state.  The main theorem below
asserts that the distributed authorization system mimics the behaviour
of the centralized reference monitor that \anSA represents.
\begin{theorem}[Safety] \label{thm-safety} 
  Suppose \aPState satisfies
  the state invariants in Appendix \ref{app-invariants}, and \TransRel{ \aPState }{ \aTransID }{ \aPState'
  }. Then the following statements hold: 
  \begin{inparaenum}
   \item If $\aTransID = \TIDRequest{ \aPerm }{ \aTicket }$ then
         $\SATrans ( \EffState{ \aPState }, \aPerm ) = \EffState{
           \aPState' }$.
   \item If $\aTransID$ is not of the form \TIDRequest{ \_ }{ \_ } then
         $\EffState{ \aPState } = \EffState{ \aPState' }$.
  \end{inparaenum}
\end{theorem}
A proof of the theorem above can be found in Appendix \ref{app-proof}.

The next theorem asserts that the client can eventually obtain a
working capability (before garbage collection occurs),
even if tickets are misplaced.
\begin{theorem}[Liveness] \label{thm-liveness}
Suppose \aPState satisfies the state
  invariants in Appendix \ref{app-invariants}.  Then there exists a (possibly empty) sequence of
  transitions
  $\aPState = \aPState_0 \TransRelArrow{ \aTransID_1 } \aPState_1
  \TransRelArrow{ \aTransID_2 } \ldots \TransRelArrow{ \aTransID_n }
  \aPState_n = \aPState'$, such that each $\aTransID_i$ is neither
  \TIDFlush nor \TIDDrop{ \_ },
  $\aPState' = \PState{ \_ }{ \_ }{ \RState{ \_ }{ \anExR' } }{
    \aCState' }$, and $\aCState'$ contains a capability \CCap{
    \TSSerial }{ \_ } for which $\TSSerial \geq \TSValid$ and
  $\TSSerial \geq \TSLast{ \anExR' }$.
\end{theorem}
A proof of liveness is given in Appendix \ref{app-proof}.


\section{Multiple Resource Servers}
\label{sec-multi}

We have been assuming that, when the client requests the authorization
server to grant access to a pool of resources, the entire pool is
guarded by a single resource server.  This section presents an
extension to HCAP for accommodating resource pools guarded by multiple
resource servers.

Every resource server has a unique identifier \RSID.  We write
$\aKey_\RSID$ to denote the shared secret established between the
resource server \RSID and the authorization server.  In \S
\ref{sec-overview}, a permission is defined to be an
operation-resource pair. We assume that the resource identifier within
a permission \aPerm also identifies the resource server $\RS(\aPerm)$ that holds the
named resource, and that it takes only $O(1)$ time
to reconstruct $\RS(\aPerm)$ from
\aPerm. (This
is true if the resource is identified by a URI, as in the
implementation reported in \S \ref{sec-performance}.)

\paragraph{New Concepts.}  
Our design of the multiple resource servers extension aims to preserve
the security guarantees of \S \ref{sec-security}. To this end, the
design is based on three concepts.

\textbf{(1) Baton holding.} A global invariant is that, at most one
resource server tracks the exception list \MapEx{ \aSessID } for a
session \aSessID. That resource server is said to be ``holding the
\Dfn{baton} (i.e., \MapEx{ \aSessID }) for session \aSessID.''  This
allows the security proofs in \S \ref{sec-security} to (mostly)
transfer to this new setting, with one exception.  Suppose a resource
server \RSID does not hold the baton for \aSessID.  Then a malicious
client may replay to \RSID an outdated capability for \aSessID.  The
resource server would not be able to differentiate between the
following two cases: Is it the case that (i) no resource server holds
the baton (i.e., the baton is garbage collected), or (ii) another
resource server holds the baton (a replay attack)?  Therefore, in the
extended HCAP scheme, the authorization server tracks an additional
boolean flag for each session to differentiate between (i) and (ii).

\textbf{(2) Remote capability validation.}  A capability \aCap for
session \aSessID is signed by a specific shared secret, say
$\aKey_{\RSID_1}$, so that only $\RSID_1$ knows how to check the
hash value of \aCap.  Capability validation also involves consulting
\MapEx{ \cdot }, and thus $\RSID_1$ needs to hold the baton for \aSessID
as well.  When \aCap is presented along an access request to a
resource server $\RSID_2$ different from $\RSID_1$, $\RSID_2$ will now
have to request $\RSID_1$ to perform capability validation on its
behalf (aka \Dfn{remote capability validation}).

To facilitate remote capability validation, a capability now has the
form \EncapCapMulti{ \aKey }{ \aUID }{ \Validator }{ \aSessID }{
  \TSSerial }{ \anSAFrag }. The new element \Validator explicitly identifies the
resource server who knows the secret key \aKey.  In other words,
$\aKey = \aKey_\Validator$, and thus \Validator can validate the
capability.

To preserve the efficiency of stationary transitions,
we further assume the SA satisfies the property below:
\begin{equation} \label{eqn-homo-trans}
\forall \anSAState, \anSAState' \in \SAStates \,.\,
\forall \aPerm, \aPerm' \in \SAAlpha \,.\,
  \SATrans (\anSAState, \aPerm) \!=\! \SATrans (\anSAState', \aPerm')
  \rightarrow \RS(\aPerm) \!=\! \RS(\aPerm')
\end{equation}
Intuitively, transitions going into a state are triggered by permissions
that can be exercised on the same resource server.  Consequently,
exercising a stationary permission never causes remote capability
validation (and baton passing, see below). This ensures stationary
transitions are always efficient. Property \eqref{eqn-homo-trans}
also makes it natural to associate a resource server $\RS(\anSAState)$
to every state \anSAState: transitions into \anSAState
can always be conducted on resource server $\RS(\anSAState)$. 

\textbf{(3) Baton passing.}  When $\RSID_2$ requests $\RSID_1$ to
perform remote capability validation, $\RSID_1$ will pass \MapEx{
  \aSessID } to $\RSID_2$ after validation succeeds. This step is
known as \Dfn{baton passing}. The intention is that stationary
transitions performed on $\RSID_2$ after that point will be efficient
(i.e., not involving remote capability validation).

\paragraph{Server States and Session Initialization.}
In addition to \MapMonitor{ \cdot }, \MapState{ \cdot }, \MapSerial{
  \cdot } and \MapFragment{ \cdot }{ \cdot }, the authorization server
maintains a boolean flag \MapBaton{ \aSessID } for each session
\aSessID.  The invariant is that \MapBaton{ \aSessID } is true if and
only if at least one of the resource servers has the baton for
\aSessID (i.e., \MapEx{ \aSessID } is defined on that resource
server).  When a new session \aSessID starts, the authorization server
sets \MapBaton{ \aSessID } to false, since \MapEx{ \aSessID } is not
yet defined on any resource server.

\paragraph{Authorization.}

When a request
$(\aUID, \aPerm, \EncapCapMulti{ \aKey }{ \aUID }{ \Validator }{
  \aSessID }{ \TSSerial }{  \anSAFrag })$ is presented to a
resource server \RSID, authorization is performed via Algorithm
\ref{algo-auth-mrs}, which is composed of three sections.

(1) The first section, consisting of line \ref{line-mrs-perm-check}
only, has no counterpart in Algorithm \ref{algo-auth}. That line
checks whether the resource server \RSID can actually exercise the
permission \aPerm. This check is necessary because \aPerm can only be
exercised by $\RS ( \aPerm )$.

(2) The second section, made up of lines
\ref{line-mrs-begin-validate}--\ref{line-mrs-end-validate}, has the
same role as lines \ref{line-integrity}--\ref{line-ex-last} in
Algorithm \ref{algo-auth}. The section validates the integrity of the
capability, and initializes \MapEx{ \aSessID } if necessary.

If \RSID is specified as the validator, then Algorithm
\ref{algo-validate} is invoked locally on \RSID to perform the
validation logic.  Algorithm \ref{algo-validate} is mostly equivalent
to lines \ref{line-integrity}--\ref{line-ex-last} in Algorithm
\ref{algo-auth}, with one exception. When \RSID does not hold the
baton (i.e., \MapEx{ \aSessID } is not defined), it is because either
(i) no resource server holds the baton, or (ii) another resource
server holds the baton. Case (ii) corresponds to a replay attack. This
is prevented by lines
\ref{line-mrs-as-check-begin}--\ref{line-mrs-as-check-end} of
Algorithm \ref{algo-validate}, which contact the authorization server
to confirm case (i).
 
If the validator \Validator is a resource server other than \RSID,
then \RSID requests \Validator to run Algorithm \ref{algo-validate}
remotely (line \ref{line-mrs-remote-validation}). (During remote
capability validation, the identifier \RSID in Algorithm
\ref{algo-validate} refers to the validator \Validator.)  If
validation succeeds, then the baton is passed to \RSID (lines
\ref{line-mrs-passing-baton}--\ref{line-mrs-receiving-baton}).

(3) The third section is composed of lines
\ref{line-mrs-transition-begin}--\ref{line-mrs-transition-end}. This
last section of Algorithm \ref{algo-auth-mrs} plays the same role as
lines \ref{line-transition-begin}--\ref{line-transition-end} in
Algorithm \ref{algo-auth}.  Two points are worth noting. First, the
capabilities and update requests issued by \RSID are signed by the
$\aKey_\RSID$ instead of $\aKey_\Validator$ (lines
\ref{line-mrs-issue-upd} \& \ref{line-mrs-issue-cap}). Second, when
\aPerm is stationary, assumption \eqref{eqn-homo-trans} guarantees
that $\RSID = \Validator$, and thus no new capabilities need to issued
(line \ref{line-mrs-issue-nothing}).

\begin{algorithm}
\KwIn{A client access request $(\aUID, \aPerm, \aCap%
)$, where
   \aUID is the client's authenticated identity,
   \aPerm is the permission to be exercised, and
   \aCap is a capability \EncapCapMulti{ ? }{ ? }{ \Validator }{ \aSessID 
    }{ \TSSerial }{ \anSAFrag }
    for session
    \aSessID and validator \Validator, 
    such that $\anSAFrag = \SAFrag{ \FDefs }{ \aCurrName }$ and
    $\FDefs ( \aCurrName ) = (\SP, \FTrans)$.
}
\KwOut{A set of tickets, or a failure response.}
\KwData{The resource server maintains the following persistent data:
  (a) its identity \RSID,
  (b) a secret $\aKey_\RSID$ it shares with the authorization server,
  and
  (c) a map \MapEx{\cdot} that assigns exceptions to 
  session IDs.
}
 \lIf{$\RS ( \aPerm ) \neq \RSID$}{%
    \Return{failure}\label{line-mrs-perm-check}%
 }
 \If{ $\Validator = \RSID$\label{line-mrs-begin-validate}}{
    Invoke Algorithm \ref{algo-validate} locally on \RSID to validate
    \aCap\label{line-mrs-local-validation}\;
    \lIf{Algorithm \ref{algo-validate} fails}{\Return{failure}}
 }
 \Else{
    Request \Validator to run Algorithm
    \ref{algo-validate} to validate \aCap\label{line-mrs-remote-validation}\;
    \If{Algorithm \ref{algo-validate} succeeds}{
      \Validator sends \RSID the contents of \MapEx{ \aSessID }\label{line-mrs-passing-baton}\;
      \Validator deletes its copy of \MapEx{ \aSessID }\label{line-mrs-validator-clean-up}\;
      \RSID stores the received contents locally in \MapEx{ \aSessID }\label{line-mrs-receiving-baton}\label{line-mrs-end-validate}\;
    }
    \lElse{\Return{failure}}
 }

    \If{\label{line-mrs-transition-begin}\label{line-mrs-stationary-first}$\aPerm \in \SP$}{
      Exercise permission \aPerm\;
      \Return{\label{line-mrs-stationary-last}\label{line-mrs-issue-nothing}$\emptyset$}\;
    }
    \ElseIf{\label{line-mrs-transitioning-first}$\aPerm \in \Dom{ \FTrans }$}{
       Exercise permission \aPerm\;
       $\aTS \leftarrow \mathit{current\_time}()$\;
       $\MapEx{ \aSessID } \leftarrow 
              \ExEx{\aPerm}{\aTS}{\MapEx{ \aSessID }}$\label{line-mrs-ex-update}\;
       $\anSAFrag' \leftarrow \FDo{ \anSAFrag }{ \aPerm }$\label{line-mrs-frag-trans}\;
       \If{$\anSAFrag' = \Unknown$}{
         \Return{\label{line-mrs-issue-upd}$\{ 
          \EncapUpdateMulti{ \aKey_\RSID }{\aUID}{\RSID}{\aSessID}{\MapEx{\aSessID}} \}$}\;
       }
       \Else{
         \Return{\label{line-mrs-transitioning-last}\label{line-mrs-issue-cap}$\{ \EncapCapMulti{ \aKey_\RSID }{\aUID}{\RSID}{\aSessID}{\aTS}{\anSAFrag'} \}$}\;
       }
    } 
    \lElse{%
       \Return{failure}\label{line-mrs-deny}\label{line-mrs-transition-end}%
    }
\caption{\label{algo-auth-mrs}Authorize access request
  (mult.~res.~servers)}
\end{algorithm}

\begin{algorithm}
\KwIn{An authenticated client identifier
   \aUID and a capability \aCap of the form \EncapCapMulti{ ? }{ ?
  }{ \Validator }{ \aSessID }{ \TSSerial }{ \anSAFrag }.
}
\KwOut{Success or failure.}
\KwData{The resource server maintains the following persistent data:
  (a) its identity \RSID,
  (b) a secret $\aKey_\RSID$ it shares with the authorization server,
  and
  (c) a map \MapEx{\cdot} that assigns exceptions to session IDs.
}
 \lIf{\label{line-mrs-integrity}\aCap is not signed by
   $\aKey_\RSID$ for \aUID
 }{%
  \Return{failure}
 }
 \If{\label{line-mrs-ex-first}\label{line-mrs-overwrite-ex}\MapEx{ \aSessID } is
   undefined}{%
   Request the authorization server to confirm that (a) $\MapBaton{
     \aSessID } = \mathit{false}$,
    and (b) $\TSSerial = \MapSerial{ \aSessID }$; on confirmation 
   \MapBaton{ \aSessID } is set to $\mathit{true}$\label{line-mrs-as-check-begin}\;
   \lIf{(a) and (b) are not confirmed}{\Return{failure}\label{line-mrs-as-check-end}}
   $\MapEx{ \aSessID } \leftarrow \ExNil{ \TSSerial }$%
 }
 \lElseIf{$\TSSerial > \TSLast{ \MapEx{ \aSessID } }$}{%
   $\MapEx{ \aSessID } \leftarrow \ExNil{ \TSSerial }$%
 }
 \lElseIf{\label{line-mrs-too-old}$\TSSerial < \TSLast{ \MapEx{ \aSessID } }$}{%
     \Return{failure}\label{line-mrs-ex-last}%
 }
 \Return{success}
\caption{\label{algo-validate}Validate capability}
\end{algorithm}

\paragraph{Baton Compression.}
Resource servers have different memory capacities, and thus when a
large baton is passed, the receiving resource server may not have
enough memory to store the baton.  This concern is addressed by an
implementation technique known as \Dfn{baton compression}, which
allows us to bound the length of batons by the size of SA fragments.

The key observation is that, when a transition is performed on an SA
fragment (Algorithm \ref{algo-auth-mrs}, line
\ref{line-mrs-frag-trans}, \FDo{ \anSAFrag }{ \aPerm }), the
underlying transition diagram (\FDefs) remains unchanged.  Thus the
transitions recorded in an exception list visit states from the same
transition diagram.  Eventually, a state will be revisited when the
length of the transition history exceeds the number of states in the
transition diagram. When this happens, the transition sequence
contains a loop.  Even if loops are eliminated from the transition
history, the authorization server can still reconstruct the current SA
state.

With the baton compression mechanism, when a transitioning permission
is exercised by a resource server for a session \aSessID, the resource
server will check that the length of the resulting exception list
\MapEx{ \aSessID } does not exceed the number of states in the SA
fragment of the capability associated with the request.  If the check
fails, then the transition history is examined for the presence of
loops. Any discovered loops are eliminated, and thus \MapEx{ \aSessID
} is ``compressed'' into a loop-free transition history with a length
bounded by the size of the SA fragment stored in the capability of the
request.  Consequently, baton passing involves only very small
payloads with sizes proportional to that of capabilities.

\paragraph{Hard and Soft Garbage Collection.}

Recall that a design objective of HCAP is to minimize communication
with the authorization server (\Obj{2}). Compared to core HCAP (\S
\ref{sec-HCAP}), the extended protocol has one additional
communication with the authorization server: line
\ref{line-mrs-as-check-begin} in Algorithm \ref{algo-validate}.  To
make this additional communication an infrequent event, we have
devised an optimization technique by enriching the garbage collection
mechanism as follows.  

When garbage collection is triggered on the resource server, two types
of garbage collection may be performed for each session \aSessID.
Session \aSessID undergoes \Dfn{hard garbage collection} when \MapEx{
  \aSessID } is reset to undefined (i.e., deallocated).  This is the
same sort of GC performed by core HCAP.  Session \aSessID undergoes
\Dfn{soft garbage collection} when \MapEx{ \aSessID } is set to
\ExNil{ \aTS }, where \aTS is the largest timestamp in the exception
list \MapEx{ \aSessID } prior to garbage collection.  Soft GC clears
the exception list without relinquishing the baton (i.e., a nil entry
is kept).  The resource server is configured in such a way that a
session with lots of activities will only undergo soft GC, and a
session that has been inactive for an extended time will undergo a
hard GC.  The effect is that batons are retained for active
sessions. Therefore, the extra communication with the authorization
server (line \ref{line-mrs-as-check-begin} of Algorithm
\ref{algo-validate}) is only performed when a session that has
remained inactive for an extended time becomes active again.
What constitutes ``extended time'' is a configurable parameter, meaning
that the extra communication with the authorization server can be
made as infrequently as possible.

As in core HCAP, the garbage-collected (whether hard or soft)
exception lists are sent to the authorization server.  Along with each
exception list, we now have to also indicate, by way of a boolean
flag, whether the baton of the corresponding session has been retained
(i.e., soft GC). The authorization server will use these boolean flags
to update its \MapBaton{ \cdot } map.

\section{Implementation and Experiments}
\label{sec-performance}

\subsection{Implementation}

We implemented the extended HCAP protocol in Java.  The implementation
is based on CoAP \cite{CoAP}, a lightweight, UDP-based variant of HTTP
commonly used in IoT environments. An HCAP permission in this context
is a pair composed of a CoAP method (e.g., GET) and an URI.  We rely
on the Californium-core library to provide CoAP functionalities.
DTLS, which provides TLS-like features for UDP, is used to enable
secure communication and mutual authentication.  X.509 certificates
are used by DTLS for authentication.  DTLS features are provided by
the Scandium-core library.

Tickets, as well as the contents of \MapEx{ \cdot } that are sent
during GC, are encoded as either JSON \cite{JSON} or CBOR \cite{CBOR}
objects.  JSON is a human readable, lightweight data interchange
format which is a subset of the JavaScript Programming Language. CBOR
is a variant of JSON that represents data in a compact binary format.
We used the jackson-dataformat-cbor and jackson-dataformat-json
libraries respectively for CBOR and JSON encoding/decoding.

\begin{figure}
\centering
\includegraphics[width=.67\linewidth]{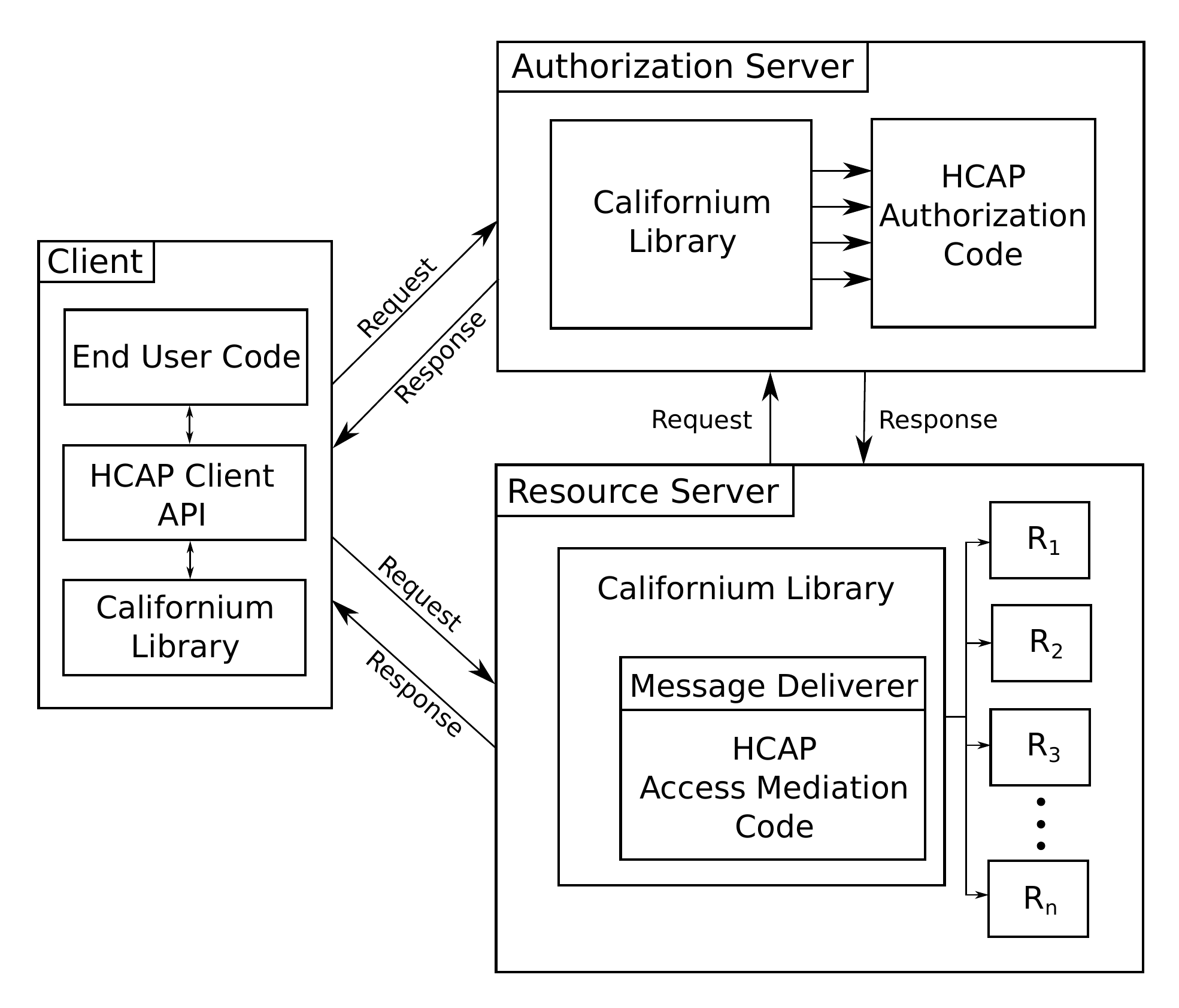}
\caption{Architecture of our implementation.\label{fig:architecture}}
\end{figure}

Our implementation consists of three reusable components
(Fig.~\ref{fig:architecture}).  The first is a client-side library
(``HCAP Client API'' in Fig.~\ref{fig:architecture}), which allows
client code to issue HCAP requests to the authorization server and the
resource servers.  The payload of an HCAP request contains both a
capability and the actual payload which the client might want to
deliver. Thus our request payload is a JSON map with two keys, mapping
to the capability and the actual payload.  The second component is a
CoAP server that acts as the authorization server (``HCAP
Authorization Code'' in Fig.~\ref{fig:architecture}).  It offers
RESTful services \cite{REST} for issuing capabilities, processing
update requests, and performing garbage collection.  The third
component allows IoT vendors to add HCAP access control
functionalities to a resource server that runs on the Californium
framework.  More specifically, we developed a \Dfn{message deliverer}
for mediating accesses.  Within the Californium framework, a message
deliverer is a hook method which intercepts a CoAP request before it
reaches the intended resource.  We developed a custom message
deliverer that checks a capability for its validity before passing the
request to the resources (``HCAP Access Mediation Code'' in
Fig.~\ref{fig:architecture}).

\begin{figure}[t]
  \centering
  \subfloat[Incomplete~SA~Fragments\label{subfig-frag}]{
    \includegraphics[width=0.4\linewidth]{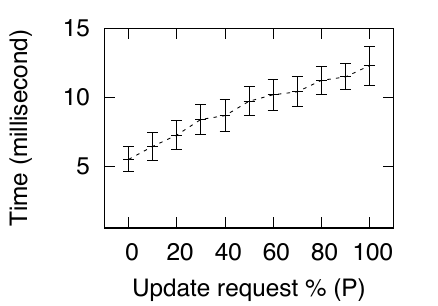}
  } 
  \qquad
  \subfloat[SA Complexity\label{subfig-comp}]{
    \includegraphics[width=0.4\linewidth]{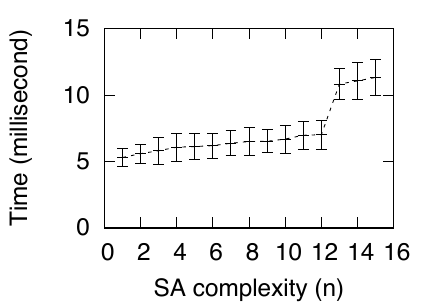}
  }

  \subfloat[Garbage Collection\label{subfig-gc}]{
    \includegraphics[width=0.4\linewidth]{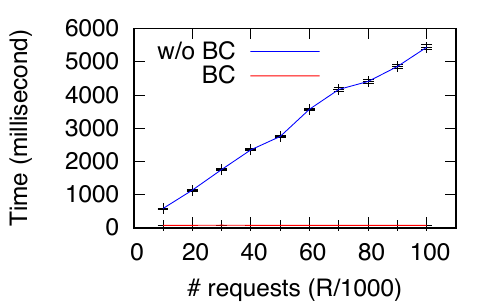}
  } 
  \qquad
  \subfloat[Baton Passing\label{subfig-bp}]{
    \includegraphics[width=0.4\linewidth]{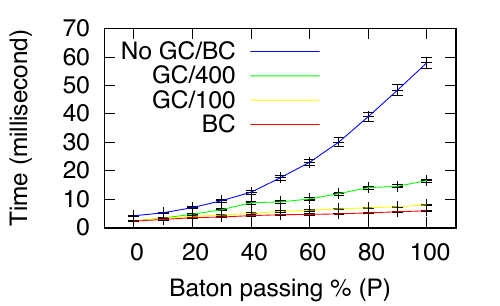}
  }
  \caption{Performance impact of various features of extended HCAP (95\%
    confidence interval).\label{fig-results}}
\end{figure}

CoAP is designed for small data transfers, but sometimes the size of
data being transferred might be too large to fit in a single packet
(e.g., during garbage collection). In this case we make use of
blockwise transfers \cite{BlockwiseTransfers}, another functionality
provided by CoAP to segment large chunks of data as blocks and send
them over.  This allows us to perform garbage collection in an
efficient manner.

\subsection{Empirical Evaluation}

We conducted four experiments, each involving a separate machine for
the client, the authorization server, and a resource server (except
for Experiment 4, which involves two resource servers). The client and
the resource servers were running on separate 3.6 Ghz Intel Core i7
(4790) machines with 8 GB of RAM, running Fedora 24 4.8 and OpenJDK
Runtime Environment (1.8). The authorization server was running on a
2.66 Ghz Intel Xeon(R)(X5355) machine with 24 GB of RAM, running
Fedora 22 4.8 and OpenJDK Runtime Environment (1.8). All machines were
connected via wired LAN on a 1 Gbps line.

The first communication between the client and the servers takes
considerable amount of time, this is due to a DTLS session being
established between the two communicating parties. To eliminate this
confounding factor, we send a dummy request (ping) to the servers
before the experiments start.  We used ``SHA256withECDSA'' to generate
signatures, and the key size was 256 bits.

\paragraph{Experiment 1: Incomplete SA Fragments.} If the SA fragment
in a capability does not provide enough information for the resource
server to construct its next capability, an update request is returned
to the client (Algorithm \ref{algo-auth-mrs}, line
\ref{line-mrs-issue-upd}), causing an extra communication with the
authorization server.  The purpose of this experiment is to assess the
performance impact of incomplete SA fragments.

The experimental protocol session involves an SA
$\anSA = \SA{\{ \aPerm_0, \aPerm_1 \}}{ \{ \anSAState_0,
  \anSAState_1 \}}{ \SAInit }{ \SATrans }$, such that
$\SATrans ( \anSAState_i , \aPerm_0 ) = \anSAState_i$,
$\SATrans ( \anSAState_i , \aPerm_1 ) = \anSAState_{1-i}$.  In short,
$\aPerm_0$ is stationary, and $\aPerm_1$ is transitioning.  Every
capability issued by the authorization server carries an SA fragment
$\anSAFrag = \SAFrag{ \FDefs }{ \aCurrName }$ for which
$\FDefs( \aCurrName ) = (\SP, \FTrans)$ and
$\Ran{\FTrans} = \{ \Unknown \}$. In other words, the resource server
returns an update request whenever the request involves $\aPerm_1$.
In each experimental configuration, we had a client generating 100
access requests, so that $P\%$ of the requests involved $\aPerm_1$.
The client would then bring any update request to the authorization
server before the next access request was generated.  We repeated this
for $P = 0, 10, \ldots, 100$.  We measured the average time it took
for the client to complete an access request, including the overhead
of contacting the authorization server in case an update request was
returned.

The results are depicted in
Fig.~\ref{fig-results}\subref{subfig-frag}.  The average request
handling time is between 5.5 to 12.3 milliseconds ($10^{-3}$ sec), an
acceptable range considering that the high-end (12.3 millisec)
corresponds to the case when every access request results in an update
request ($P = 100\%$).


\paragraph{Experiment 2: SA Complexity.}
The purpose of this experiment is to assess the performance impact of
embedding a complex SA in a capability.  The experimental protocol
sessions involve SA of the following form:
$\anSA_n = \SA{ \SAAlpha_n }{ \SAStates_n }{ \SAInit }{ \SATrans_n }$,
where $\SAAlpha_n = \{ \aPerm_0, \ldots, \aPerm_{n-1} \}$,
$\SAStates_n = \{ \anSAState_0, \ldots, \anSAState_{n-1} \}$, and
$\SATrans_n ( \anSAState_i , \aPerm_j ) = \anSAState_j$. In short, there
is a transition between every ordered pair of states in $\anSA_n$.
The SA fragments embedded in capabilities incorporate the \emph{full}
specification of the SA.  We varied $n$ from 1 up to 15.  For each
value of $n$, a client issued 100 randomly generated access requests
to the resource server. We measured the average time for the
client to complete one request.

The results are depicted in
Fig.~\ref{fig-results}\subref{subfig-comp}.  As $n$ increases, the
number of transitions in $\anSA_n$ grows quadratically, so does the
size of the SA fragment.  When the value of $n$ increased from 12 to
13, we see a sudden jump in the request handling time.  This is
because, when $n \leq 12$, the entire SA fragment can be fitted into a
single UDP packet, but multiple packets are needed when $n > 13$. That
was when CoAP blockwise transfer kicked in. This highlights the
advantage of keeping SA fragments moderate in size.

\paragraph{Experiment 3: Garbage Collection.}
The purpose of this experiment is to evaluate the performance impact
of garbage collection (GC).  Reusing the SA $\anSA_{12}$ from
Experiment 2, we deployed 100 clients to issue transitioning requests
to the resource server.  As each transitioning request for session
\aSessID was served, a new exception entry was added to \MapEx{
  \aSessID }.  Once all the requests were issued, GC was triggered,
and the entire contents of \MapEx{ \cdot } were transfered to the
authorization server for updating \MapState{ \cdot }.  We measured the
time for GC to complete.  Let $R$ be the total number of requests made
by the clients right before GC was triggered.  The experiment was
repeated for $R$ $=$ 10,000, 20,000, \ldots, 100,000. In addition, for
each $R$, the experiment was repeated 100 times to obtain the average
GC overhead.  These experiments were conducted in two experimental
configurations: (1) \textbf{w/o BC}, in which baton compression was
turned off, and (2) \textbf{BC}, in which baton compression was turned
on.  This allowed us to observe how GC interacted with baton
compression

The results are depicted in Fig.~\ref{fig-results}\subref{subfig-gc},
with the horizontal axis corresponding to $R$ divided by 1,000.
Without baton compression (\textbf{w/o BC}), the size of \MapEx{ \cdot
} grew in proportion to $R$ (i.e., the number of transitioning
requests issued to the resource server), and GC time grew accordingly.
If we amortize GC time over individual requests, then the per-request
GC overhead ranges between 54 and 58 microseconds ($10^{-6}$ sec),
that is, between 0.78\% and 0.83\% of the average request handling
time (Experiment 2, $n=12$).

When baton compression was turned on (\textbf{BC}), the overhead of
garbage collection was significantly reduced. This was because the
length of an exception list \MapEx{ \aSessID } is bounded by 12 (i.e.,
the number of states in $M_{12}$). Thus the total size of \MapEx{
  \cdot } is never above $12 \times 100 = 1,200$.  Amortizing the GC
overhead over individual requests, the per-request GC overhead ranges
between 0.72 and 7.5 microseconds ($10^{-6}$ sec), that is, between
$0.01\%$ and $0.10\%$ of the average request handling time.

\paragraph{Experiment 4: Baton Passing.}

The purpose of this experiment is to evaluate the performance impact
of baton passing. Two resource servers were involved. The SA $M_2$
from Experiment 2 was reused, so that the two resource servers played
the role of $\RS(\anSAState_0)$ and $\RS(\anSAState_1)$ respectively.
Consequently, baton passing was triggered when and only when a
transitioning permission was exercised.  A single client was
configured to issue a total of 1,000 requests to the resource servers,
so that $P\%$ of the requests involved baton passing. The experiment
was repeated for $P = 0, 10, \ldots, 100$. The average time required
to complete one authorization request was recorded for each $P$.  In
addition, the experiments were conducted in four different
experimental configurations, so that we could observe how the overhead
of baton passing was affected by garbage collection and baton
compression:
\begin{asparaenum}
\item \textbf{No GC/BC.} Both garbage collection and baton compression were
  turned off.
\item \textbf{GC/400.} Garbage collection was triggered after every 400
  transitioning requests, but baton compression was turned off.
\item \textbf{GC/100.} This configuration is similar to GC/400 except that
  garbage collected was triggered after every 100 requests.
\item \textbf{BC.} Baton compression was turned on, but garbage collection was
  turned off.
\end{asparaenum}

The results are depicted in
Fig.~\ref{fig-results}\subref{subfig-bp}. In the case of \textbf{No
  GC/BC}, the baton sizes (i.e., lengths of exception lists) grew
indefinitely because neither garbage collection nor baton compression
was turned on. Therefore, baton passing incurred significant overhead
(approximately 60 milliseconds per authorization request when
$P=100\%$).  Turning on garbage collection (\textbf{GC/400} and
\textbf{GC/100}) significantly reduced the overhead of baton passing,
because a baton was reduced to empty every time it was garbage
collected, and thus batons were never given the chance to grow too
long. We also notice that the more frequently garbage collection was
triggered (e.g., more frequently in \textbf{GC/100} than in
\textbf{GC/400}), the overhead of baton passing became smaller.  The
most promising result, however, is that of \textbf{BC}, in which baton
compression was turned on (even without the help of GC): it took only
6 milliseconds to complete an authorization request even when
$P=100\%$.  This is because the batons (i.e., exception lists) are
kept to a size of 2 ($M_2$ has only two states).  In this case, baton
passing involved only the sending of a single UDP packet.

\section{Conclusion and Future Work}
We argued that the physical embeddedness and process awareness of IoT
devices impose a natural sequencing of accesses, which can be
exploited for realizing Least Privilege.  To this end, we proposed
HCAP, a distributed capability system for enforcing history-based
access control policies in a decentralized manner.  We formally
established the security guarantees of HCAP and empirically
demonstrated that the performance of HCAP is competitive.

The following are some directions for future work:
\begin{inparaenum}[(1)]
\item integrating HCAP into UMA or OpenID,
\item adding fault tolerance into HCAP,
\item compilation of workflow specification and/or UMP specification
  \cite{Becker-Nanz:2010} into SA fragments for use in HCAP, and
\item incorporating context awareness (e.g., time, location, sensor
  inputs) into HCAP.
\end{inparaenum}

\section*{Acknowledgment}

This work is supported in part by an NSERC Discovery Grant and a
Canada Research Chair.


\appendix

\section{State Invariants}
\label{app-invariants}

Let $\aPState \in \PStates{ \anSA }$ be a protocol state of the form
\PState{ \TSClock }{ \anAState }{ \anRState }{ \aCState }, where
$\anAState = \AState{ \anSAState }{ \TSState}$, and
$\anRState = \RState{ \TSValid }{ \anExR }$.  We articulate below
state invariants for a protocol session.
{\renewcommand{\labelenumi}{\Inv{\theenumi}}
\begin{enumerate}
\item 
  \label{inv-first}
  \label{inv-global-clock} The global clock \TSClock is larger than
  all the timestamps (e.g., \TSState, \TSValid, \TSSerial,
   etc) within the protocol state. (That is, \TSClock is
  always a fresh timestamp.)

\item \label{inv-ex-time} 
  Let 
   $\anExR = \ExEx{ \aPerm_m }{ \aTS_m }{
     \ExEx{ \aPerm_{m-1} }{ \aTS_{m-1} }{
       \ldots 
       \ExEx{ \aPerm_1 }{ \aTS_1 }{ \ExNil{ \aTS_0 }}
     }
   }$ for some $m \geq 0$. (That is, 
   $\anExR = \ExNil{ \aTS_0 }$ if 
   $m = 0$.)
   Then $\aTS_0 < \aTS_1 < \ldots < \aTS_m$.



\item \label{inv-major-cases} One of the following two cases holds:
  (a) $\TSLast{ \anExR } < \TSState$; (b)
  $\TSFirst{ \anExR } = \TSState$, $\TSFirst{ \anExR } \geq \TSValid$,
  and $\FDoEx{ \anSAFrag_{\anSA, \anSAStateA} }{ \anExR }$ is defined.  In case (a)
  above, there are two further subcases: (i) $\anExR = \ExNil{ \aTS }$
  for some $\aTS < \TSValid$, and $\TSValid \leq \TSState$; (ii) 
  $\TSFirst{ \anExR } \geq \TSValid$.

  \textbf{Remarks:} Intuitively, case (a) corresponds to situations
  when the authorization server has up-to-date information about the
  current state of the security automaton, while case (b) corresponds
  to situations when the knowledge of the authorization server is
  lagging behind that of the resource server. Under case (a), subcase
  (i) arises when garbage collection occurs, and subcase (ii) occurs
  right after an update request is presented to the authorization
  server.

\item \label{inv-as-cap} If $\TSLast{ \anExR } < \TSState$ (i.e.,
  case (a) of \Inv{\ref{inv-major-cases}}), then for every
  $\CCap{ \TSSerial }{ \anSAFrag } \in \aCState$, one of the three
  cases below holds: (i)
  $\TSSerial < \TSLast{ \anExR }$; (ii) $\TSSerial < \TSValid$; (iii)
  $\TSSerial = \TSState$ and $\anSAFrag = \anSAFrag_{\anSA, \anSAStateA}$.

\item \label{inv-rs-cap} If $\TSFirst{ \anExR } = \TSState$ (i.e., case (b) of
  \Inv{\ref{inv-major-cases}}), then for every
  $\CCap{ \TSSerial }{ \anSAFrag } \in \aCState$, one of the following
  two cases holds: (i) $\TSSerial < \TSFirst{ \anExR }$; (ii) 
  $\TSSerial \in \Times{ \anExR }$, and \FDoExUpto{ \TSSerial }{
    \anSAFrag_{\anSA, \anSAStateA} }{ \anExR } is an SA fragment
  (i.e., not \Unknown) identical to \anSAFrag.

\item \label{inv-as-upd}
  If $\TSLast{ \anExR } < \TSState$ (i.e.,
  case (a) of \Inv{\ref{inv-major-cases}}), then for every
  $\CUpdate{ \anEx } \in \aCState$, $\TSState > \TSLast{ \anEx }$.

\item \label{inv-rs-upd} 
  If $\TSFirst{ \anExR } = \TSState$ (i.e., case (b) of
  \Inv{\ref{inv-major-cases}}), then for every 
   $\CUpdate{ \anEx } \in \aCState$, one of the following two cases
   holds: (i) $\TSLast{ \anEx } < \TSFirst{ \anExR }$; (ii)
   $\anEx = \anExR$ and $\FDoEx{ \anSAFrag_{\anSA, \anSAStateA} }{ \anExR } = \Unknown$.

\label{inv-last}

\end{enumerate}
}
The following proposition can be established.
\begin{proposition}[State Invariants]
  The initial state $\aPState_0$ satisfies conditions \Inv{\ref{inv-first}} to
  \Inv{\ref{inv-last}}. In addition,
  if \aPState satisfies conditions \Inv{\ref{inv-first}} to
  \Inv{\ref{inv-last}}, and \TransRel{ \aPState }{ \aTransID }{
    \aPState' }, then $\aPState'$ also satisfies those conditions.
\end{proposition}

\section{Proof of Theorems}
\label{app-proof}

\paragraph{Proof of Theorem \ref{thm-safety} (Safety).}
We prove statement (1) first. Here,
$\aTransID = \TIDRequest{ \aPerm }{ \aTicket }$.  By the preconditions
of \TRRequestStationary and \TRRequestTransitioning,
$\aTicket \in \aCState$ and
$\aTicket = \CCap{ \TSSerial }{ \anSAFrag }$.  We demonstrate below
that
$\EffState{ \aPState' } = \SATrans ( \EffState{ \aPState }, \aPerm )$.
By \Inv{\ref{inv-major-cases}}, one of the following two
cases holds.

\textbf{Case 1:} $\TSLast{ \anExR } < \TSState$. In this case,
$\EffState{ \aPState } = \anSAStateA$. According to
\Inv{\ref{inv-as-cap}}, the only way for the preconditions of
\TRRequestStationary and \TRRequestTransitioning to be satisfied is
when $\TSSerial = \TSState$ and
$\anSAFrag = \anSAFrag_{\anSA, \anSAStateA}$. Note that
$\anSAFrag_{\anSA, \anSAStateA}$ is by construction safe for \anSA in
\anSAStateA, in other words, \EffState{ \aPState }. Since the effects
of \TRRequestStationary and \TRRequestTransitioning ensure that
$\FDo{ \anSAFrag }{ \aPerm }$ is defined, Lemma \ref{lemma-fragment}
ensures that $\SATrans( \EffState{ \aPState }, \aPerm )$ is also
defined.  \textbf{Subcase 1.1:} \TRRequestStationary is responsible
for the transition. Since \aPerm is stationary for \anSAFrag, Lemma
\ref{lemma-fragment} implies that
$\SATrans ( \anSAStateA, \aPerm ) = \anSAStateA$. In addition,
\TRRequestStationary causes $\anSAStateA' = \anSAStateA$,
$\anExR' = \ExNil{\TSSerial}$, and $\TSState' = \TSFirst{ \anExR'
}$. Therefore,
$\EffState{ \aPState' } = \anSAStateA = \SATrans ( \anSAStateA, \aPerm
) = \SATrans ( \EffState{ \aPState }, \aPerm )$.  \textbf{Subcase
  1.2:} \TRRequestTransitioning is responsible for the transition.
Then \TRRequestTransitioning causes $\anSAStateA' = \anSAStateA$,
$\anExR' = \ExEx{ \aPerm }{ \TSClock }{ \ExNil{ \TSSerial } }$, and
$\TSState' = \TSFirst{ \anExR' }$. Therefore,
$\EffState{ \aPState' } = \SATransEx ( \anSAStateA, \anExR' ) =
\SATrans ( \anSAStateA, \aPerm ) = \SATrans ( \EffState{ \aPState },
\aPerm )$.

\textbf{Case 2:} $\TSFirst{ \anExR } = \TSState$.  In this case,
$\EffState{ \aPState } = \SATransEx ( \anSAStateA, \anExR )$.
According to \Inv{\ref{inv-rs-cap}}, the only way for the
preconditions of \TRRequestStationary and \TRRequestTransitioning to
be satisfied is when $\TSSerial = \TSLast{ \anExR }$ and
$\anSAFrag = \FDoEx{ \anSAFrag_{ \anSA, \anSAStateA } }{ \anExR }$.
Since $\anSAFrag_{ \anSA, \anSAStateA }$ is by construction safe for
\anSA in \anSAStateA, Lemma \ref{lemma-fragment} guarantees that
\anSAFrag is safe for \anSA in \EffState{ \aPState }.  The effects of
\TRRequestStationary and \TRRequestTransitioning ensure that
$\FDo{ \anSAFrag }{ \aPerm }$ is defined, and thus by Lemma
\ref{lemma-fragment}, $\SATrans( \EffState{ \aPState }, \aPerm )$ is
also defined.  \textbf{Subcase 2.1:} \TRRequestStationary is
responsible for the transition. Then $\anExR' = \anExR$,
$\anSAStateA' = \anSAStateA$, and $\TSState' = \TSState$. Lemma
\ref{lemma-fragment} guarantees that
$\SATrans ( \EffState{ \aPState }, \aPerm ) = \EffState{ \aPState }$.
Thus
$\EffState{ \aPState' } = \SATransEx ( \anSAStateA', \anExR' ) =
\SATransEx ( \anSAStateA, \anExR ) = \SATrans ( \SATransEx (
\anSAStateA, \anExR ), \aPerm ) = \SATrans ( \EffState{ \aPState },
\aPerm )$.  \textbf{Subcase 2.2:} \TRRequestTransitioning is
responsible for the transition. Then
$\anExR' = \ExEx{ \aPerm }{ \TSClock }{ \anExR }$,
$\anSAStateA' = \anSAStateA$, $\TSState' = \TSState$.  Consequently,
$\EffState{ \aPState' } = \SATransEx ( \anSAStateA', \anExR' ) =
\SATrans ( \SATransEx ( \anSAStateA, \anExR ), \aPerm ) = \SATrans (
\EffState{ \aPState }, \aPerm )$.

We now turn to prove statement (2). Here, \aTransID is not of the form
\TIDRequest{ \_ }{ \_ }. None of \TRIssue, \TRRecover and \TRDrop
causes a change of the effective SA state.  Thus we consider only
\TRFlush and \TRUpdate.  For \TRFlush, the effective SA state may
change only if $\TSState = \TSFirst{ \anExR }$ and
$\TSState' > \TSLast{ \anExR' }$. In that case,
$\EffState{ \aPState } = \SATransEx ( \anSAStateA, \anExR ) =
\EffState{ \aPState' }$.  For \TRUpdate,
$\EffState{ \aPState' } = \SATransEx ( \anSAStateA, \anEx ) =
\SATransEx ( \anSAStateA, \anExR ) = \EffState{ \aPState }$ by
\Inv{\ref{inv-rs-upd}}.

\paragraph{Proof of Theorem \ref{thm-liveness} (Liveness).}
By \Inv{\ref{inv-major-cases}}, there are two cases:

\textbf{Case 1:} $\TSLast{ \anExR } < \TSState$.  According to
\TRIssue, the transition \TIDIssue produces for the client a
capability with a serial number \TSState, which is greater than
\TSLast{ \anExR } (according to the case assumption).  That $\TSState
\geq \TSValid$ is guaranteed by \Inv{\ref{inv-major-cases}}.

\textbf{Case 2:} $\TSFirst{ \anExR } = \TSState$. The transition
\TIDIssue is first executed, producing for the client a capability
with a serial number \TSState, which is equal to \TSFirst{ \anExR }
(by the case assumption).  Then the transition \TIDRecover{ \aTicket } is
executed.  \Inv{\ref{inv-rs-cap} } ensures that \TRRecover produces
for the client either a capability \CCap{ \TSLast{ \anExR } }{
  \anSAFrag' } or an update request \CUpdate{ \anExR }.  If a
capability is produced, then the serial number is \TSLast{ \anExR
}. Such a serial number is not less than \TSValid (by
\Inv{\ref{inv-major-cases}}).  If an update request is produced, then
executing the transition \TIDUpdate{ \CUpdate{ \anExR } } will put us
back into \textbf{Case 1}.

\end{document}